\documentclass[11pt,a4paper]{article}
\usepackage{amsmath,amssymb,bbold,epsfig}
\usepackage{amsfonts}
\usepackage{slashed}
\usepackage{dsfont}
\usepackage{cite}
\usepackage{graphicx,color}
\usepackage{bbm}
\usepackage[normalem]{ulem}
\usepackage[usenames,dvipsnames]{xcolor}
\definecolor{lightgray}{gray}{0.90}
\usepackage[colorlinks=true,linkcolor=red,citecolor=green,urlcolor=cyan,
bookmarks=true,bookmarksopen=true,pdfpagemode=None,pdfstartview=FitH]{hyperref}
\newcommand{\appsection}{\addtocounter{section}{1}\setcounter{equation}{0}
                         \renewcommand{\thesection}{\Alph{section}}
}
\renewcommand{\theequation}{\arabic{equation}}
\newcommand{\be}{\begin{equation}}
\newcommand{\ee}{\end{equation}}
\newcommand{\bea}{\begin{eqnarray}}
\newcommand{\eea}{\end{eqnarray}}

\DeclareMathOperator{\diag}{diag}
\begin{document}

\title{\vspace{-2cm}
\vglue 1.3cm
\Large \bf
Collective neutrino oscillations and neutrino wave packets}
\author{
{Evgeny~Akhmedov,\!$^1$\footnote{Also at the National Research Centre
Kurchatov Institute, Moscow, Russia}~\thanks{Email: \tt
akhmedov@mpi-hd.mpg.de}
~\;Joachim Kopp$^{2}$\thanks{Email: \tt jkopp@uni-mainz.de}~ and
\,Manfred Lindner$^1$\thanks{Email: \tt lindner@mpi-hd.mpg.de}
\vspace*{3.5mm}
} \\
{\normalsize\em
$^1$Max-Planck-Institut f\"ur Kernphysik, Saupfercheckweg 1, \vspace*{-1mm}}\\
{\normalsize\em
69117 Heidelberg, Germany
\vspace*{0.15cm}}
\\
{\normalsize\em
$^2$PRISMA Cluster of Excellence and Mainz Institute for
Theoretical Physics, \vspace*{-1mm}} \\
{\normalsize\em
 Johannes Gutenberg University, 55099 Mainz, Germany
\vspace*{0.15cm}}
}
\date{}

\maketitle
\thispagestyle{empty}
\vspace{-0.8cm}
\begin{abstract}
Effects of decoherence by wave packet separation on collective neutrino
oscillations in dense neutrino gases are considered. We estimate the length of
the wave packets of neutrinos produced in core collapse supernovae and the
expected neutrino coherence length, and then proceed to consider the
decoherence effects within the density matrix formalism of neutrino flavour
transitions.  First, we demonstrate that for neutrino oscillations in vacuum
the decoherence effects are described by a damping term in the equation of
motion of the density matrix of a neutrino as a whole (as contrasted to that of
the fixed-momentum components of the neutrino density matrix). Next, we
consider neutrino oscillations in ordinary matter and dense neutrino
backgrounds, both in the adiabatic and non-adiabatic regimes. In the latter
case we study two specific models of adiabaticity violation -- one with
short-term and another with extended non-adiabaticity. It is demonstrated that,
while in the adiabatic case a damping term is present in the equation of motion
of the neutrino density matrix (just like in the vacuum oscillation case), no
such term in general appears in the non-adiabatic regime.
\end{abstract}
\vspace{0.2cm}
\vspace{0.3cm}

\newpage

\section{\label{sec:intro}Introduction}

Neutrino oscillations in dense neutrino environments existing 
in the early Universe and at certain stages of supernova explosions may differ
significantly from oscillations in vacuum or in ordinary matter.  In dense
neutrino backgrounds, coherent neutrino--neutrino interactions may strongly
affect the flavour evolution of the neutrino system, leading to a variety of
collective oscillation phenomena.  These include synchronized oscillations
\cite{synch1,synch2,Pantaleone:1992eq, synch3,synch4,synch5,synch6,RaffTamb},
bipolar oscillations \cite{bipolar,synch4,duan1,Duan:2006an,
hann1,Fogli:2007bk,duan2}, spectral splits and swaps
\cite{splits1,splits2,splits3,splits4} and multiple spectral splits
\cite{multisplits}.  These phenomena have attracted a great deal of attention
recently, see Refs.~\cite{review1,review2} for reviews and extensive lists of
literature.

Since in supernovae and in the early Universe neutrinos are produced at
very high densities, their production processes are characterized by a high
degree of space-time localization and, as a result, the wave packets (WPs)
describing the states of produced neutrinos are very short in configuration
space.  In particular, in \cite{Kersten,KerstSmir} the estimate $\sigma_x\sim
10^{-11}$~cm was obtained for the spatial length of supernova neutrino WPs by
considering the properties of the WPs of the electrons involved in the neutrino
production.  In Section \ref{sec:size1} of this paper we arrive at a similar
result, basing our estimate on different considerations.  
The WPs of different propagation eigenstates%
\footnote{\label{foot:1}
By propagation eigenstates we mean the eigenstates of the effective neutrino 
Hamiltonian, which in general includes neutrino potentials coming from 
neutrino coherent forward scattering on the particles of the usual matter 
and on background neutrinos.}
composing a given produced flavour-eigenstate neutrino in general propagate
with different group velocities. Therefore, very short neutrino WPs suggest
that the packets of the individual propagation eigenstates would quickly
separate in space, leading to decoherence and hence suppression of neutrino
oscillations.  Simple estimates of the coherence length of supernova neutrinos,
i.e.\ the distance over which the WPs of their propagation eigenstate
components fully separate, yield $L_{\rm coh} \sim 10$ km \cite{Kersten}. One
might therefore expect that decoherence by WP separation would significantly
affect flavour transitions of supernova neutrinos, and, in particular,
collective neutrino oscillations, which are expected to occur at a distance
$\sim 100$~km from the supernova's center. Numerical simulations show, however,
that in reality the situation is more complex.

Decoherence effects on collective neutrino oscillations are very little studied,
with most existing work focusing on understanding the evolution
of neutrinos in the coherent regime. To our knowledge, the only papers that
directly address the conditions under which coherence is lost
are Refs.~\cite{RaffTamb} and \cite{AM}.%
\footnote{Decoherence effects on flavour transformations of supernova
neutrinos were also studied in Ref.~\cite{KerstSmir}. However,
the influence of the coherence loss on collective neutrino oscillations
has not be considered there.}
In \cite{RaffTamb} it has been demonstrated that in a dense uniform and
isotropic neutrino gas decoherence due to late-time dephasing of different
neutrino momentum modes does not occur provided that the neutrino number
density is high enough. As a result, synchronized neutrino oscillations that
can take place in such a system are not destroyed by decoherence.  In
\cite{RaffTamb} an interpretation of this fact in terms of mode locking due to
neutrino self-interaction was suggested; however, the WP picture of neutrino
flavour transitions was not explicitly discussed.  In Ref.~\cite{AM} possible
decoherence effects in dense neutrino gases (or their absence) were studied
from the point of view of spatial separation of neutrino WPs, both in the
adiabatic and non-adiabatic regimes of neutrino flavour evolution.  The results
of \cite{RaffTamb} were confirmed from this standpoint; in addition, some {\em
ad hoc} assumptions used in the analytic approach developed in \cite{RaffTamb}
were shown to be redundant.

In the present paper we continue studies of the effects of
decoherence by WP separation on collective neutrino oscillations in dense
neutrino gases. Our goal is in particular to develop a space-time picture
of such decoherence effects, complementary to the more widely used
momentum space approach, in which the neutrino ensemble is split
into individual modes of fixed momentum.
We address the question of whether WP separation
can be described by effective evolution equations for the density matrix of a
neutrino WP as a whole (rather than of its components for a given momentum or
energy). We derive such effective equations and discuss in detail their domain
of applicability.%
\footnote{The present paper supersedes and replaces our previous paper on this
subject \cite{AKL3} (currently withdrawn), in which the domain of validity
of this effective evolution equation was overestimated.}

We would like to emphasize here that our configuration-space 
approach to decoherence is effectively equivalent to the usual one based on the 
momentum-space considerations, provided that the momentum distributions of the  
individual neutrino states are properly taken into account. However, 
as we shall see, it allows new insights into the space-time development of 
decoherence. 

The paper is organized as follows. In Section~\ref{sec:size} we estimate the
spatial length of the WPs of neutrinos produced in core collapse supernovae
and discuss the neutrino coherence length with respect to WP separation. In
Section~\ref{sec:densmat} we derive and discuss equations of motion (EoMs)
satisfied
by the neutrino density matrix in
dense matter and neutrino backgrounds. We start by reviewing the evolution
equation for the neutrino state vector (Section~\ref{sec:stvec}), the
neutrino density matrix in flavour space (Section~\ref{sec:dm}) and
the density matrix describing neutrino oscillations in vacuum for
neutrinos described by Gaussian WPs (Section~\ref{sec:vac}). In
Section~\ref{sec:vac} we also derive the EoMs satisfied by the
coordinate-averaged and un-averaged neutrino density matrices in vacuum
and show that these equations essentially coincide, provided that in
the latter case the total time derivative is understood as the Liouville
operator. Moreover, we demonstrate that the evolution equation of the
density matrix of a neutrino as a whole (as contrasted to the equations for
fixed-momentum components of the density matrix), as well as that of the
coordinate-space density matrix, explicitly contains a damping term
which describes decoherence by WP separation.
In Section~\ref{sec:medium} we consider the
EoM for the neutrino density matrix in normal matter
and dense neutrino backgrounds, both in the adiabatic (Section~\ref{sec:adiab})
and in the non-adiabatic (Section~\ref{sec:nonadiab}) regimes. For the latter,
we consider two specific models of adiabaticity violation in
Sections~\ref{sec:model1} and~\ref{sec:model2}.
We show that, just like for vacuum oscillations, the EoM of the neutrino
density matrix in the adiabatic regime contains a damping term which leads to
decoherence by WP separation at sufficiently late times. However, no
such term in general appears in the non-adiabatic regime.
We summarize and discuss our results in Section~\ref{sec:disc}. In the Appendix
we briefly review the flavour spin formalism for description of neutrino
oscillations in dense neutrino backgrounds and derive within this
formalism the expression for the non-adiabaticity parameter in the case of
2-flavour neutrino oscillations. We also give there an approximate analytic
solution of the EoM for neutrino flavour spin vectors in the case of
synchronized neutrino oscillations.

\section{\label{sec:size}Supernova neutrino wave packets and the coherence
length}

An important parameter characterizing decoherence phenomena in neutrino
propagation is the coherence length $L_{\rm coh}$. This is the distance
traveled by neutrinos beyond which the WPs corresponding to different
propagation eigenstates composing the produced neutrino flavour eigenstate
separate by more than their spatial length $\sigma_x$.
Once this has happened, the neutrino WPs no longer overlap, and their coherence
is lost.%
\footnote{
Decoherence by WP separation can in principle be undone by neutrino
detection if the individual detection processes are characterized by a
sufficiently high energy resolution (i.e.\ are sufficiently
delocalized in space and time) \cite{Kiers}. However, at least for terrestrial
detection of neutrinos from a galactic supernova this
would require unrealistically high energy resolution,
and so we disregard this possibility.}
To find the coherence length of supernova neutrinos, let us first estimate
$\sigma_x$.

\subsection{\label{sec:size1}The length of supernova neutrino WPs}

In Refs.~\cite{Kersten,KerstSmir} the length $\sigma_x$ of supernova neutrino
WPs was estimated assuming that it is dominated by the length of the WPs of the
charged leptons (mostly electrons and positrons) participating in the neutrino
production process. However, as we show below, while this assumption is
correct in the case of neutrino production in $e^+ e^-$  annihilation, it is
actually not valid when neutrinos are produced in processes with participation
of nucleons.

It has been demonstrated in \cite{Beuthe,parad} that the length of the
neutrino WP is determined by the spatial and temporal localization of the
neutrino production process and is typically dominated by the latter. The
temporal localization of neutrino production (i.e.\ the duration of the
production process) is given by the overlap time $\sigma_t$ of the WPs of all
particles taking part in the process.
In ref.~\cite{Beuthe} it has been shown
that this time can usually
be estimated as
\be
\sigma_t\sim
\frac{\sigma_{x2}}{|\vec{v}_2-\vec{v}_1|}\,,
\label{eq:sigmat}
\ee
where $\vec{v}_1$ and $\vec{v}_2$ are the velocities of the particles with
the shortest and next-to-shortest WPs in configuration space, and
$\sigma_{x2}$ is the spatial length of the next-to-shortest WP.
This result has a very simple interpretation: since
the production process requires the overlap
of the WPs of all participating particles,
it is over as soon as at least one of these WPs ceases to overlap with all
the others. The time scale of neutrino production is therefore
typically given by the interval during which the shortest WP ``slides'' along 
the second shortest one, which leads to the estimate in eq.~(\ref{eq:sigmat}).
Note, however, that this result may need a modification if there
is a longer WP which moves much faster than the shortest and the
second shortest ones. We will discuss this point in more detail later on.

For neutrino production processes with participation of nucleons, the shortest
and the second shortest WPs are those of the involved nucleons, and their 
lengths can be estimated as $\sigma_{x1}\sim \sigma_{x2}\lesssim r_0$, where
$r_0$ is the average distance between the nucleons in the medium. The relative
velocities of the nucleons are of the order of their mean thermal velocity
$\bar{v}$.
The parameters $r_0$ and $\bar{v}$ can be estimated from the relations
\be
\frac{m \bar{v}^2}{2} =\frac{3}{2}T\,, \qquad n_{bar}\approx \frac{\rho}{m}
\approx
\big(\tfrac{4}{3}\pi r_0^3\big)^{-1},
\label{eq:vr01}
\ee
where $\rho$, $n_{bar}$ and $T$ are the matter density, baryon number
density and temperature
at the neutrino production point, respectively, and $m$ is the nucleon mass.
This gives
\be
\bar{v}\approx 0.179\,\Big(\frac{T}{10\;\rm MeV}\Big)^{1/2}\,,\qquad
r_0\approx 7.36\times 10^{-13}~{\rm cm}\,\Big(\frac{\rho}{10^{12}\;\rm g/cm^3}
\Big)^{-1/3}\,.
\label{eq:vr02}
\ee
For the length $\sigma_x$ of the neutrino WP we then find
\be
\sigma_x\,\simeq\, v_g\sigma_t\,\lesssim\, 4.1\times 10^{-12}~\rm{cm}\,
\Big(\frac{\rho}{10^{12}\;\rm g/cm^3}\Big)^{-1/3}
\Big(\frac{T}{10\;\rm MeV}\Big)^{-1/2}\,,
\label{eq:sigmax}
\ee
where $v_g\simeq 1$ is the mean group velocity of neutrino propagation
eigenstates. At density $\rho\sim 10^{12}$~g/cm$^3$ and temperature $T\sim 10$~MeV,
typical for the neutrino production region in a supernova, we have
\be
\sigma_x\lesssim 4.1\times
10^{-12}~{\rm cm}\,.
\label{eq:sigmax2}
\ee
Note that $\sigma_x$ depends only rather weakly on the matter density at
neutrino production: for neutrinos produced close to the electron
neutrino sphere, where $\rho\sim 5\times 10^{10}$ g/cm$^3$ \cite{JH},
$\sigma_x$ is only about a factor of 2.7 larger than our estimate
(\ref{eq:sigmax2}).

Eq.~(\ref{eq:sigmat}), on which the above estimates are based, has been obtained
in \cite{Beuthe} from a more general (but less transparent) relation under
the assumptions that the velocities $\vec{v}_1$ and $\vec{v}_2$
are not nearly equal, and that
the magnitudes of the velocities of other particles taking part in
neutrino production are not vastly different from $v_1$ and $v_2$. The latter
condition is actually not satisfied for neutrino production in supernovae, as
electrons are relativistic at temperatures $T\sim 10$ MeV, while nucleons are
non-relativistic.
Let us therefore consider the effect of the electron WP
on the duration of the neutrino production process.

It has been demonstrated in \cite{Kersten} that the length of the WPs of
electrons participating in neutrino production in supernovae can be estimated
as $\sigma_{xe}\simeq (4\pi\alpha^2 n_e)^{-1/3}$, where $\alpha$ is the fine
structure constant and $n_e$ is the electron number density. For $\rho\sim
10^{12}$ g/cm$^3$ and the electron fraction per baryon $Y_e\sim 1/2$, this
gives $\sigma_{xe}\sim 10^{-11}$ cm, which is much larger than the lengths of
the nucleon WPs. However, since electrons move much faster than nucleons, it is
conceivable that the overlap time of the electron and nucleon WPs may be
shorter than the nucleon--nucleon overlap time.  The duration of the neutrino
production process, and thus the length of the neutrino WP, would then be 
determined by the electron--nucleon overlap time.  To see if this is indeed 
the case, let us compare electron--nucleon and  nucleon--nucleon overlap 
times. For the ratio of these quantities we have
\be
\frac{\sigma_{xe}}{r_0/|\vec{v}_2-\vec{v}_1|} \sim
\frac{(4\pi\alpha^2 n_e)^{-1/3}}{[(4\pi/3)n_{bar})]^{-1/3}}
\Big(\frac{3T}{m}\Big)^{1/2}
\approx 4.2 \times \Big(\frac{1}{2Y_e}\Big)^{1/3}
\Big(\frac{T}{10\;{\rm MeV}}\Big)^{1/2}.
\label{eq:ratio1}
\ee
Thus, the overlap time of electron and nucleon WPs is typically larger than the
nucleon-nucleon overlap time by about a factor of four or more.  This means
that the effect of the electron WPs on the temporal localization of the
neutrino production processes with participation of nucleons (and therefore on
the length of the WPs of produced neutrinos) is small. For Gaussian WPs its
relative contribution to the length of neutrino WPs is suppressed by a factor
\,$\lesssim e^{-8} \sim 3\times10^{-4}$. Neutrino interactions on the way from
their birthplace in the supernova core to the neutrinosphere (from which they
essentially stream out freely) do not increase the length of the neutrino WPs.

Interestingly, even though the considerations on which our estimate of
$\sigma_x$ was based were quite different from those employed by Kersten and
Smirnov in~\cite{Kersten,KerstSmir}, numerically our result is rather close to
theirs, $\sigma_x\sim 10^{-11}$ cm.

\subsection{\label{sec:coher}Coherence length of supernova neutrinos}

Let us now turn to the coherence length of supernova neutrinos $L_{\rm coh}$.
The propagation eigenstates $\nu_i$ composing the initially produced flavour
eigenstate neutrino
in general move with different group velocities $v_{gi}$; as a result, after
having traveled a distance
\be
L_{\rm coh}\simeq \sigma_x \frac{v_g}{\Delta v_g}\,,
\label{eq:Lcoh01}
\ee
their WPs separate by a distance exceeding their length $\sigma_x$ and
therefore cease to overlap.  To find the coherence length $L_{\rm coh}$, one
therefore needs to know, in addition to $\sigma_x$, the difference of their
group velocities $\Delta v_g$. The group velocities are well defined for the
propagation eigenstates and are given by the derivatives of their energies with
respect to momentum. Therefore,
\be
\Delta v_g \,=\,\frac{\partial}{\partial p}\Delta E\,,
\label{eq:deltaVg1}
\ee
where $\Delta E$ is the difference of the energies of the corresponding
neutrino propagation eigenstates, i.e.\ the difference of the eigenvalues of
the effective neutrino Hamiltonian. For relativistic neutrinos in vacuum
(free propagation) this gives $\Delta v_g\simeq \Delta m^2/(2p^2)$,
where $p$ is the
neutrino momentum, and eq.~(\ref{eq:Lcoh01}) yields
\be
L_{\rm coh}\simeq \frac{2p^2}{\Delta m^2}
\sigma_x\,.
\label{eq:Lcoh02}
\ee

It should be stressed that in a system with more than two neutrino flavours 
there is a separate coherence length $L_{\rm coh}^{jk}$ for each pair of
propagation eigenstates $\nu_j$, $\nu_k$.  It is the distance beyond which 
the WPs of $\nu_j$ and $\nu_k$ cease to overlap. 
Similarly, the quantities $\Delta v_g$, $\Delta E$ and $\Delta 
m^2$ in eqs.~(\ref{eq:Lcoh01}), (\ref{eq:deltaVg1}) and (\ref{eq:Lcoh02}) 
should also carry indices $jk$. 
We have suppressed these indices to simplify the notation. Complete
decoherence occurs
when the distance $L$ traveled by neutrinos satisfies
$L>\max\{L_{\rm coh}^{jk}\}$ for all pairs $\{j,k\}$;
however, partial decoherence may also be of interest.

For neutrino oscillations in a medium, one has to take into account that the
effective neutrino Hamiltonian depends on the properties of the medium.
Consider for definiteness the 2-flavour case, in which the effective
Hamiltonian is a $2\times 2$ Hermitian matrix
\be
{\cal H}=\left(\begin{array}{cc}
{\cal H}_{11} & {\cal H}_{12} \\
{\cal H}_{12}^* & {\cal H}_{22}
\end{array} \right).
\label{eq:H1a}
\ee
Its eigenvalues are
\be
E_{1,2}=\frac{{\cal H}_{11}+{\cal H}_{22}}{2}\,\pm\,
\sqrt{\frac{({\cal H}_{11}-{\cal H}_{22})^2}{4}+|{\cal H}_{12}|^2} \;.
\label{eq:eigen1}
\ee
In this case
\be
\Delta v_g \,=\,\frac{\partial}{\partial p} \bigg(
2\sqrt{\frac{({\cal H}_{11}-{\cal H}_{22})^2}{4}+|{\cal H}_{12}|^2}\,\bigg).
\label{eq:deltaVg2}
\ee
In particular, for neutrino oscillations in ordinary matter, one finds from the
standard expression for the neutrino Hamiltonian ${\cal H}$ 
that~\cite{MikhSm2} 
\be
\frac{\Delta v_g}{v_g} \simeq \frac{\Delta m^2}{2p^2}\cdot
\frac{\frac{\Delta m^2}{2p}-\sqrt{2}G_F n_e c_{20}}
{\sqrt{\big(\frac{\Delta m^2}{2p}c_{20}-\sqrt{2}G_F n_e\big)^2 +
\big(\frac{\Delta m^2}{2p} s_{20} \big)^2}}\,,
\label{eq:deltaVg4}
\ee
where $G_F$ is the Fermi constant,
$s_{20}=\sin 2\theta_0$, and $c_{20}=\cos 2\theta_0$, with $\theta_0$ being the
leptonic mixing angle in vacuum. It is interesting to note that, when
the density is such that $\sqrt{2}G_F n_e c_{20}=
\Delta m^2 / (2p)$, the quantity $\Delta v_g/v_g$ vanishes.
For small values of the vacuum mixing angle $\theta_0$, the 
density required to satisfy this condition is very close to the one at the 
Mikheyev--Smirnov--Wolfenstein (MSW) resonance, where $\sqrt{2}G_F n_e = 
c_{20} \Delta m^2 / (2p)$. Note
that for small $\theta_0$, the density interval where $\Delta v_g/v_g$ is
suppressed is very narrow; outside this region the absolute value of the last
factor on the right hand side of eq.~(\ref{eq:deltaVg4}) is of order one,
i.e.\ $|\Delta v_g|/v_g$ is of the same order of magnitude as
$\Delta m^2/(2p^2)$.

Consider next the case of 2-flavour oscillations in a dense 
uniform and isotropic neutrino gas. In this case it is convenient to study 
neutrino flavour evolution in the framework of the ``flavour spin'' formalism 
\cite{SR}. In this picture, for each neutrino mode $\omega=\Delta m^2/(2p)$,
the traceless part of the neutrino density matrix in flavour space
$\rho(t,\vec{p})$ is written as a linear combination of the Pauli matrices.
The coefficients of the Pauli matrices form a flavour spin vector
$\vec{P}_\omega$, whose evolution is a precession around the (in general
time-dependent) Hamiltonian vector $\vec{H}_\omega$.
The latter, in turn, is formed by the coefficients of the expansion of the
traceless part of the neutrino Hamiltonian in terms of the Pauli matrices
(see the Appendix). The vectors
$\vec{H}_\omega$ depend on the global flavour spin vector $\vec{P}$, which
is the sum of all the individual $\vec{P}_\omega$. It is easy to show that
in the 2-flavour case the difference of the eigenvalues of the effective
neutrino Hamiltonian is $\Delta E=|\vec{H}_\omega|$, so that
\be
\Delta v_g \,=\,\frac{\partial}{\partial p} |\vec{H}_\omega|\,.
\label{eq:deltaVg3}
\ee
Substituting here the expression for $|\vec{H}_\omega|$ from
eq.~(\ref{eq:length}) yields
\be
\Delta v_g \,=\,\frac{\partial \omega}{\partial p}
\Big(
\frac{\partial}{\partial\omega}|\vec{H}_\omega|\Big)\,=
\;\frac{\Delta m^2}{2p^2}\cdot
\frac{\omega -\mu P_0c_{20}}{\sqrt{(\omega-\mu P_0c_{20})^2+
\mu^2 (P^2-c_{20}^2 P_0^2)}}\,,
\label{eq:deltaVg5}
\ee
where
$\mu=\sqrt{2}G_F n_\nu$,
$P=|\vec{P}|$, and $P_0$ is defined as $P(t=0)$.
Note that the global flavour spin vector $\vec{P}$ is in general time
dependent, and so are $\Delta v_g$ and $L_{\rm coh}$. We did not indicate the
time dependence in eq.~(\ref{eq:deltaVg5}) explicitly in order not to
overload the notation.
Just as in the case of oscillations in ordinary matter, the absolute value of
the last factor on the right hand side of eq.~(\ref{eq:deltaVg5}) is typically
of order unity except in a relatively narrow region around the point in the
parameter space where $\omega -\mu P_0c_{20}=0$.%
\footnote{Indeed, the second term under the square root in
eq.~(\ref{eq:deltaVg5}) vanishes for $\mu\le\mu_0$ (where $\mu_0$ is the
threshold value of $\mu$ discussed in the Appendix) because in this case
$P\to c_{20}P_0$; thus the absolute value of the last factor in
eq.~(\ref{eq:deltaVg5}) is essentially equal to one.
For $\mu\gg \mu_0$ we have $P\to P_0$ and the the last factor in
eq.~(\ref{eq:deltaVg5}) is $\simeq -c_{20}$. For intermediate values
of $\mu$ the modulus of this factor is also of order unity.
}

Thus,
in most of the parameter space a simple estimate $\Delta v_g/v_g\simeq
\Delta m^2/(2p^2)$ (and therefore the expression for $L_{\rm coh}$ in
eq.~(\ref{eq:Lcoh02})) should be quite accurate, i.e.\ the quantities
$\Delta v_g$ and $L_{\rm coh}$ are to a good approximation time independent.

It should be emphasized that the discussion of the coherence length in this
section, strictly speaking, applies only to the oscillations in vacuum and
to adiabatic flavour evolution in medium. In the non-adiabatic regime
the very notion of the coherence length may lose its sense; we shall discuss
this issue in more detail in Sections~\ref{sec:model2} and~\ref{sec:disc}.

\section{\label{sec:densmat}Equation of motion for neutrino density matrix}

\subsection{\label{sec:stvec}Evolution of the neutrino state vector}

We start with considering the evolution of an individual neutrino in a
background of other neutrinos and ordinary matter.
The coordinate-space neutrino
state vector
$|\nu(t,\vec{x})\rangle$ can be Fourier-expanded as
\be
|\nu(t,\vec{x})\rangle=\int\frac{d^3 p}{(2\pi)^3}\,
e^{i\vec{p}\vec{x}}\,
|\nu(t,\vec{p})\rangle\,,
\label{eq:wp1m}
\ee
where
$|\nu(t,\vec{p})\rangle$ is the time-dependent state
vector for a neutrino of a given momentum $\vec{p}$. We will assume that
$|\nu(t,\vec{p})\rangle$
satisfies the standard Schr\"odinger-like evolution equation
\be
i\frac{d}{dt}|\nu(t,\vec{p})\rangle = {\cal H}_{\vec{p}}(t)
|\nu(t,\vec{p})\rangle\,,
\label{eq:Schr1}
\ee
In the following we will often be using the matrix representation, where 
${\cal H}_{\vec{p}}(t)$ is an $N_f\times N_f$ Hermitian matrix with $N_f$ 
being the number of neutrino flavours.  Then in the flavour basis \cite{SR}
\be
{\cal H}_{\vec{p}}(t)=U_0 \frac{M^2}{2p}U_0^\dag + \hat{V}(t) +
\sqrt{2}G_F\int\frac{d^3q}{(2\pi)^3}[\rho(t,\vec{q})-\bar{\rho}(t,\vec{q})]
(1-\vec{v}_{\vec{p}}\vec{v}_{\vec{q}})\,.
\label{eq:H1}
\ee
Here $M$ is the diagonal matrix of neutrino masses, $U_0$ is the leptonic
mixing matrix in vacuum, $\hat{V}(t)$ is the matrix of potentials due to 
coherent forward scattering of neutrinos on ordinary matter, $G_F$ is the 
Fermi constant, $\rho(t,\vec{p})$ and $\bar{\rho}(t,\vec{p})$ are the density
matrices of, respectively, neutrinos and antineutrinos of momentum $\vec{p}$
(to be discussed in more detail below), and $\vec{v}_{\vec{p}}$ is the velocity
of neutrinos with momentum $\vec{p}$.  The last term on the right hand side of
eq.~(\ref{eq:H1}) is due to coherent forward scattering of the test neutrino on
background neutrinos.  Note that the first term in~(\ref{eq:H1}) depends only
on $p=|\vec{p}|$. In isotropic matter and neutrino backgrounds, the same is
true for the whole Hamiltonian; in particular, the last term in (\ref{eq:H1})
reduces to $\sqrt{2}G_F[\rho(t)-\bar{\rho}(t)]$, where
\be
\rho(t)=
\int\frac{d^3q}{(2\pi)^3}\rho(t,\vec{q}) = \int\frac{4\pi q^2dq}{(2\pi)^3}
\rho(t,|\vec{q}|),
\label{eq:rho1}
\ee
and similarly for $\bar{\rho}(t)$.

Let us introduce the neutrino propagation eigenstates and leptonic mixing
matrix in medium. At any time $t$ the Hamiltonian ${\cal H}_{\vec{p}}(t)$ can
be diagonalized by a unitary transformation
\be
{\cal H}_{\vec{p}}(t) = U(t) {\cal H}_{\vec{p}}^{d}(t) U^\dag(t)\,,
\label{eq:H2m}
\ee
where $U(t)$ is the leptonic mixing matrix in medium
and ${\cal H}_{\vec{p}}^{d}(t)$ is the diagonal matrix of the
eigenvalues $E_j(t,\vec{p})$ of ${\cal H}_{\vec{p}}(t)$, i.e.\
\be
{\cal H}_{\vec{p}}^{d}(t)=\diag\big(E_1(t,\vec{p}), E_2(t,\vec{p}),...\big)\,.
\label{eq:eigen}
\ee
Because of the time dependence of ${\cal H}_{\vec{p}}(t)$, its eigenstates and
eigenvalues are also $t$-dependent; at a given $t$ they are called
instantaneous eigenstates and eigenvalues.%
\footnote{Since the Hamiltonian ${\cal H}_{\vec{p}}(t)$ depends on the density
matrices $\rho$ and $\bar{\rho}$, which themselves must be obtained from
the evolution equations based on eq.~(\ref{eq:Schr1}), the instantaneous
eigenstates and eigenvalues of ${\cal H}_{\vec{p}}(t)$ are not known {\it a
priori}, unlike in the case of neutrino oscillations in ordinary matter.
For our discussion, however, it is sufficient to know that such eigenstates
and eigenvalues exist. }
At any fixed instant of time the neutrino flavour states
$|\nu_\alpha(\vec{p})\rangle$ can be represented as linear superpositions of
the propagation eigenstates $|\nu_j(t, \vec{p})\rangle$:
\be
|\nu_\alpha(\vec{p})\rangle=\sum_j U^*_{\alpha j}(t)
|\nu_j(t,\vec{p})\rangle.
\label{eq:expan1a}
\ee
Since the mixing matrix in medium $U(t)$ diagonalizes the $\vec{p}$-dependent
Hamiltonian ${\cal H}_{\vec{p}}(t)$, it is also $\vec{p}$-dependent; we do not
indicate this dependence explicitly to simplify the notation.

Let a neutrino be produced at a time $t_0$ in the flavour state $\alpha$
($\alpha = e, \mu$ or $\tau$).
According to eq.~(\ref{eq:expan1a}), its state vector can be written as a
linear combination of the state vectors of the propagation eigenstates, so that
\be
|\nu(t_0,\vec{p})\rangle=
|\nu_\alpha(\vec{p})\rangle=\sum_j U^*_{\alpha j}(t_0)
|\nu_j(t_0,\vec{p})\rangle.
\vspace*{-2.5mm}
\label{eq:expan2}
\ee
We will be assuming that the propagation eigenstates $\nu_j$ composing the
initially produced flavour state are described by WPs with momentum-space
wave functions (momentum distribution amplitudes)
$f_{\vec{p}_j}(\vec{p})$, where $\vec{p}_j$ is the centroid of the
momentum distribution.
For each $\vec{p}$-component of the WP one can then write
\be
|\nu_j(t_0,\vec{p})\rangle= f_{\vec{p}_j}(\vec{p})|\nu_j^{(0)}(t_0,\vec{p})
\rangle,
\label{eq:wpmom}
\ee
where $|\nu_j^{(0)}(t,\vec{p})\rangle$ are the state vectors of the propagation
eigenstates
satisfying
\be
\langle \nu_j^{(0)}(t,\vec{p})|\nu_k^{(0)}(t,\vec{p}\,')\rangle =
(2\pi)^3\delta^{3}(\vec{p}-\vec{p}\,')\delta_{jk}\,.
\label{eq:norm0}
\ee
For the amplitudes $f_{\vec{p}_j}(\vec{p})$  we adopt the normalization condition
\be
\int\frac{d^3 p}{(2\pi)^3}\,
|f_{\vec{p}_j}(\vec{p})|^2=1\,.
\label{eq:norm1}
\ee
The momentum distribution is characterized by its
effective width $\sigma_p$. An example we will often be using below is the
Gaussian WP:
\be
f_{\vec{p}_j}(\vec{p})=
(2\pi/\sigma_p^2)^{3/4}\,
\exp\Big[-\frac{(\vec{p}-\vec{p}_j)^2}{4\sigma_p^2}\Big].
\label{eq:gauss1m}
\vspace*{1.0mm}
\ee
The plane-wave limit is recovered for $\sigma_p\to 0$, which yields 
$f_{\vec{p}_j}(\vec{p})=[(2\pi)^3/\sqrt{V}]\delta^3(\vec{p}-\vec{p}_j)$,
where $V$ is the normalization volume.

Combining eqs.~(\ref{eq:expan2}) and (\ref{eq:wpmom}), we find for the initial
state
\be
|\nu(t_0,\vec{p})\rangle=\sum_j U^*_{\alpha j}(t_0)
f_{\vec{p}_j}(\vec{p})|\nu_j^{(0)}t_0,\vec{p}\rangle.
\vspace*{-2.5mm}
\label{eq:expan3}
\ee
The neutrino state vector $|\nu(t,\vec{p})\rangle$ at $t>t_0$ can then be found
by solving the EoM (\ref{eq:Schr1}) with the initial
condition (\ref{eq:expan3}). The coordinate space neutrino state vector in the
WP picture is then obtained from eq.~(\ref{eq:wp1m}).

Let us now return to the neutrino flavour evolution equation.
In non-uniform matter and neutrino
backgrounds, the potential $\hat{V}$ and the neutrino density matrix $\rho$
should obviously depend also on the neutrino coordinate; then how can the
coordinate-independent eqs.~(\ref{eq:Schr1}) and (\ref{eq:H1}) be obtained?
In deriving these equations, neutrinos are implicitly assumed to be essentially
pointlike objects. For a pointlike neutrino moving along a classical trajectory
$\vec{x}=\vec{x}(t)$ its coordinate is fully determined by its
propagation time, and one can consider $\hat{V}$ and $\rho$ as
functions of $t$ only, rather than of $(t, \vec{x})$.
In other words, what matters is only the values of
$\hat{V}$, $\rho$ and $\bar{\rho}$ at the point reached by the neutrino at
time $t$.

The question remains, of course, as to how this can be reconciled with the
fact that eqs.~(\ref{eq:Schr1}) and (\ref{eq:H1}) are supposed to describe
flavour evolution of neutrinos of definite momenta, which correspond to
plane waves rather than to pointlike states. The answer is that these
`plane waves' should be considered as the limit of wave packets (WPs) when
their momentum spread $\sigma_p$, though still small compared to the momentum
$p$, is nevertheless large compared to the inverse of the distances over
which $\hat{V}$, $\rho$ and $\bar{\rho}$ vary significantly.

Let us discuss this point in more detail.
Eqs.~(\ref{eq:Schr1}) and (\ref{eq:H1}) (as well as the corresponding
equations in the case of the usual MSW effect \cite{W,MS}) are usually
obtained under the assumptions that neutrinos are relativistic and that the
potentials they feel are small compared to the neutrino energy and vary little
over spatial distances of the order of the neutrino de Broglie
wavelength $1/p$ (see, e.g., \cite{cardMSW,AW}).
When neutrinos are described
by WPs of finite length, one has to add the conditions that $\hat{V}$ and $\rho$
vary little also over distances of the order of the spatial size of the
neutrino WP $\sigma_x$ and over time scales of the order of the neutrino
passage time $\sim \sigma_x/v_\nu$. Since usually $\sigma_x \sim 1/\sigma_p
\gg 1/p$, these conditions are more stringent than the standard ones;
still, for the potential $\hat{V}$ they are satisfied very well for all cases
of practical interest. As for the density matrices $\rho$ and $\bar{\rho}$
entering into eq.~(\ref{eq:H1}),
in the case when neutrinos are described by WPs, their definition must include
averaging over spatial volumes whose linear size
is large compared to the
length of the neutrino WP \cite{pant1,card1}. Under this condition they also
vary little over distances $\sim \sigma_x$ and times $\sim \sigma_x/v_\nu$.
Once the above conditions are satisfied, neutrinos can be described
by finite-length WPs (for which $\vec{x}$ and $t$ are independent variables),
while their individual momentum components satisfy the EoM in
eqs.~(\ref{eq:Schr1}) and (\ref{eq:H1}).

\subsection{\label{sec:dm}Neutrino density matrix}

A convenient quantity for discussing neutrino flavour evolution in dense media
is the neutrino density operator (density matrix) in flavour space. The
1-particle density operator describing flavour transitions
of an individual neutrino in a medium consisting in general of both ordinary
matter and background neutrinos is defined as
\be
\hat{\rho}^a(t, \vec{x})=|\nu^a(t, \vec{x})\rangle\langle \nu^a(t, \vec{x})|\,,
\label{eq:rho2}
\ee
where the evolution of the state vector $|\nu^a(t,\vec{x})\rangle$ is described
by eqs.~(\ref{eq:wp1m})--(\ref{eq:H1}).
We have introduced here the superscript $a$ which labels the individual
neutrinos
($a=1,...,N$, where $N$ is the total number of neutrinos in the system).
The density operator of the whole neutrino system,
whose Fourier transform enters into the
definition of the effective Hamiltonian ${\cal H}_{\vec{p}}(t)$ in
eq.~(\ref{eq:H1}),
is the sum of the 1-particle ones:%
\footnote{The one-particle density operators $\hat{\rho}^a(t,\vec{x})$, as well 
as the relation in eq.~(\ref{eq:rho9}), are well defined provided that 
dynamical multiparticle correlations in the system can be neglected. The latter 
become sizeable only at extremely high densities, at which the mean distance 
between the neutrinos in the system is comparable with the range of the weak 
interactions. Such densities are far above those present in the supernovae. 
}
\be
\hat{\rho}(t, \vec{x})=\sum_a \hat{\rho}^a(t, \vec{x})=
\sum_a|\nu^a(t, \vec{x})\rangle\langle \nu^a(t, \vec{x})|\,.
\label{eq:rho9}
\ee
The density matrix $\hat{\rho}(t, \vec{x})$ along with the 
evolution equation (\ref{eq:Schr1}) and eqs.~(\ref{eq:wp1m}) and~(\ref{eq:H1}) 
describe the flavor content of the neutrino ensemble in the mean field 
approximation. If the overall neutrino spectrum 
correctly takes into account the spectra of the individual neutrino 
WPs, both the standard approach, in which the system is treated as an ensemble 
of neutrinos of well defined energies, and the wave packet approach result in 
the same $\hat{\rho}(t, \vec{x})$ and therefore describe the same physics.

The coordinate dependence of the neutrino density operator $\hat{\rho}
(t,\vec{x})$ has two sources. First, the density operator of each individual
neutrino has a non-trivial $\vec{x}$-dependence because it corresponds to a
finite-size WP. This type of coordinate dependence disappears when one averages
the density operator over spatial volumes of a linear size $\bar{r}\gg
\sigma_x$ around each point $\vec{x}$.  ($\bar{r}$ should, however, be small
compared to the distances over which the macroscopic characteristics of the
medium change significantly).  We shall refer to this type of averaging as
coarse graining. The second source of $\vec{x}$-dependence is related to the
fact that the properties of the medium are in general coordinate dependent, and
in particular different regions of space may be characterized by different
concentration and different flavour composition of neutrinos.  This type of
coordinate dependence, which is the only one that survives upon coarse
graining, is, however, absent in the case of uniform matter and neutrino
backgrounds.

Likewise, also the momentum dependence of the neutrino density matrix has two
sources. The first one is related to the fact that each individual neutrino is
described by a WP characterized by a certain
momentum distribution amplitude. The density matrix of the neutrino as a whole
is obtained by integrating over its individual momentum modes. The resulting
1-particle density matrix will depend on the mean momentum of the
neutrino state, which is given by
the weighted average of the centroids of the momentum distributions of the
propagation eigenstates composing the produced favour state:
\be
\langle\vec{p}\rangle = \sum_j\int\frac{d^3p}{(2\pi)^3}
\,|U_{\alpha j}(t_0)|^2\, |f_{\vec{p}_j}(\vec{p})|^2\, \vec{p}\,.
\label{eq:mean2}
\ee
The individual neutrino state vector
(\ref{eq:wp1m})
and the corresponding
expression for the density matrix given in eq.~(\ref{eq:rho2})
actually depend on this mean
momentum. Therefore the summation over the index $a$ should include the
summation (integration) over the spectrum of the mean momenta
$\langle\vec{p}\rangle_a$.

What characterizes the state of the $a$th neutrino?
Given the shape of its momentum-space wave function, each neutrino is
characterized, in addition to the mean momentum discussed above, by
its production time $t_0^a$, production coordinate $\vec{x}_0^a$, and
initial flavour $\alpha$, as well as by the evolution time $t$. The latter no
longer uniquely characterizes the state of the $a$th neutrino,
which depends also on the initial conditions, which are different for
different neutrinos. In non-uniform matter and neutrino backgrounds, the
state vectors of different neutrinos will depend on their individual histories
and will no longer be given by the solutions of the same evolution equation
(\ref{eq:Schr1}) -- at a given time $t$ the Hamiltonians
${\cal H}_{\vec{p}}(t)$ will be different for different neutrinos.

The situation is much simpler in the case of isotropic and uniform
neutrino backgrounds. Isotropy means that the velocity terms in
eq.~(\ref{eq:H1}) can be omitted and eq.~(\ref{eq:rho1}) can be
utilized. Uniformity means that it is sufficient to consider the
coordinate-independent EoM (\ref{eq:Schr1}), which for a given neutrino
production time $t_0^a$ fully determines the individual neutrino state.
If all neutrinos are produced at the same time $t_0$ (as
it is often assumed in toy models), then the density operators of the
individual neutrinos are identical, and the
density operator of the complete
neutrino system is just given by the total neutrino number $N$ times the
1-particle density operator.
In reality, different neutrinos are usually produced at different times
$t_0^a$, and therefore averaging of the 1-particle density operators over
the production time has to be performed. We will discuss this issue in
Section~\ref{sec:model1}.

We shall now address the question if and when EoMs
describing the evolution of the density operators of the individual neutrinos
as wholes (rather than of their specific momentum components) can be found.
Our goal, in particular, is to examine if such EoMs would exhibit terms
responsible for the wave packet separation, thus facilitating studies of
decoherence effects. We start with the simplest case of neutrino oscillations
in vacuum.

\subsection{\label{sec:vac}Neutrino oscillations in vacuum}

For oscillations in vacuum, the Hamiltonian of the neutrino system
is given by just the first, time-independent, term on the right hand side of
eq.~({\ref{eq:H1}). The mixing matrix $U$ in eq.~(\ref{eq:expan2}) then
coincides with the leptonic mixing matrix in vacuum $U_0$, and the
propagation eigenstates $\nu_j$ coincide with the neutrino mass eigenstates.
In this subsection we will deal only with 1-particle density matrices
which are fully adequate to neutrino oscillations in vacuum.

Let a flavour eigenstate $\nu_\alpha$ be produced at $t_0=0$ and
$\vec{x}_0=0$.%
\footnote{
By this we mean that the centroid of the neutrino WP is located at
the production time $t_0=0$ at the point with the coordinate $\vec{x}_0=0$.
Here and below we neglect the finite duration of the neutrino production
process, which is important for the formation of the neutrino WP (see
Section~\ref{sec:size}), but, once its properties are defined, is irrelevant
for its subsequent evolution, which we are mostly interested in.}
The neutrino state vector at $(t, \vec{x})$ can then be written as
\be
|\nu(t,\vec{x})\rangle=\sum_j (U_0)_{\alpha j}^* \psi_j(t, \vec{x})|\nu_j
\rangle \,,
\vspace*{-2mm}
\label{eq:psi}
\ee
where $\psi_j(t, \vec{x})$ is the coordinate-space wave function (wave packet)
of the $j$th neutrino mass eigenstate. From eq.~(\ref{eq:rho2}) it then
follows that for neutrino oscillations in vacuum the 1-particle neutrino
density matrix in the mass-eigenstate basis can be written as
\be
\rho_{jk}(t, \vec{x})=(U_0)_{\alpha j}^* (U_0)_{\alpha k}^{}\,
\psi_j(t, \vec{x})\psi_k^*(t, \vec{x})\,.
\label{eq:rho2a}
\ee
We shall now assume for definiteness that the mass-eigenstate neutrinos
constituting the produced flavour state are described in momentum space by 
Gaussian WPs of width $\sigma_p$, see eq.~(\ref{eq:gauss1m}). Expanding the 
neutrino energies $E_j(p)=(p^2+m_j^2)^{1/2}$ about the peak momenta $p_j$ and 
retaining only the first two terms in the expansion (which amounts to 
neglecting WP spreading effects),%
\footnote{
Effects of spreading of WPs are strongly suppressed for ultra-relativistic 
neutrinos. Although they may have observable consequences for neutrinos 
propagating over astrophysical distances, they can be safely neglected 
for neutrino oscillations in supernovae, and in any case they do not 
affect decoherence issues \cite{KerstSmir}.}
one finds that the coordinate space WPs also have Gaussian form.
For the density matrix in the mass eigenstate basis this yields
\cite{Giunti1}
\be
\rho_{jk}(t, \vec{x})=
\frac{(U_0)_{\alpha j}^* (U_0)_{\alpha k}^{}}{(2\pi\sigma_x^2)^{3/2}}\,
\exp\Big[{-i(E_j-E_k)t+i(\vec{p}_j-\vec{p}_k)\vec{x}}
-\frac{(\vec{x}-\vec{v}_j t)^2}{4\sigma_x^2}-\frac{(\vec{x}-\vec{v}_k t)^2}
{4\sigma_x^2}\Big],
\label{eq:gauss3}
\ee
where
\be
\sigma_x\equiv 1/(2\sigma_p)\,,\qquad
E_j\equiv E_j(p_j)\,, \qquad
\vec{v}_j\equiv\left. [\partial E_j(p)/
\partial \vec{p}]\right|_{\vec{p}=\vec{p}_j},
\label{eq:def}
\ee
and similarly for $E_k$, $\vec{v}_k$. Note that $\vec{v}_j$ and $\vec{v}_k$
are the group velocities of the $j$th and $k$th neutrino mass eigenstates,
respectively.

In the realistic situations we are interested in, the density matrix is
averaged over spatial volumes of a linear size $\bar{r}$ that is small
compared to the distances over which macroscopic characteristics of the system
change significantly, but large compared to the size $\sigma_x$ of the
individual neutrino WPs. Since the WP amplitude decreases very quickly
at distances much greater than $\sigma_x$ from their center, one can
formally extend the integration over the coordinate to infinity and consider
the quantity
\be
\rho_{jk}(t) \equiv \int d^3x\, \rho_{jk}(t, \vec{x})\,.
\label{eq:rho3}\
\ee
For the Gaussian WPs from eq.~(\ref{eq:gauss3}) we find
\be
\rho_{jk}(t)=(U_0)_{\alpha j}^* (U_0)_{\alpha k}^{}
\exp\Big[{-\frac{(\vec{p}_j-\vec{p}_k)^2}{8\sigma_p^2}\Big]
\exp\Big[-
i(E_j-E_k)t+i\vec{v}_g(\vec{p}_j-\vec{p}_k)t}
-\frac{(\vec{v}_j-\vec{v}_k)^2 t^2}
{8\sigma_x^2}\Big],
\label{eq:rho4}
\ee
where $\vec{v}_g \equiv \frac{\vec{v}_j+\vec{v}_k}{2}$ is the average
group velocity of the states $\nu_j$ and $\nu_k$.
We did not indicate explicitly the $j,k$ dependence of $\vec{v}_g$ because for
relativistic neutrinos this dependence is small and because $v_g$ enters into
eq.~(\ref{eq:rho4}) multiplied by $|\vec{p}_j-\vec{p}_k|$, which is small.
Indeed, coherent production of flavour eigenstates is only possible for
$|\vec{p}_j-\vec{p}_k|\ll \sigma_p$. Keeping the $j,k$ dependence of
$\vec{v}_g$ would therefore introduce terms
$\ll \!(\Delta m^2/2E^2)\sigma_p t$.
At the same time, we keep
the last term in the exponent in eq.~(\ref{eq:rho4}),
which is of the order of
$(\Delta m^2/2E^2)^2\sigma_p^2\,t^2$
because it can become important at sufficiently large times.

Next, we notice that for ultra-relativistic neutrinos
\be
\Delta E\,
\simeq \,
\frac{\partial E}{\partial \vec{p}}\,\Delta\vec{p}\,+\,\frac{\partial E}
{\partial m^2}\,\Delta m^2\,=\,\vec{v}_g\Delta \vec{p}\,+\,
\frac{\Delta m^2}{2E},
\label{eq:expand2}
\ee
where $\Delta E\equiv E_j-E_k$, $\Delta \vec{p}\equiv \vec{p}_j-\vec{p}_k$,
$\Delta m^2=m_j^2-m_k^2$, and $E$ is the average energy of $\nu_j$ and
$\nu_k$. Introducing the notation
\be
\tilde{E}_j = E_j-\vec{v}_g \vec{p}_j\,,
\label{eq:not1}
\ee
we can rewrite eq.~(\ref{eq:rho4}) as
\be
\rho_{jk}(t) = (U_0)_{\alpha j}^* (U_0)_{\alpha k}^{}
\exp\Big[-\frac{(\vec{p}_j-\vec{p}_k)^2}{8\sigma_p^2}\Big]
\exp\Big[-i(\tilde{E}_j-\tilde{E}_k)t
-\frac{(\vec{v}_j-\vec{v}_k)^2 t^2}
{8\sigma_x^2}\Big].
\label{eq:rho5}
\ee
Differentiating this expression, we find the EoM satisfied by $\rho_{jk}(t)$:
\be
\frac{d}{dt}\rho_{jk}(t) = \Big[-i(\tilde{E}_j-\tilde{E}_k)-\frac{2t}
{L_{\rm coh}^2}\Big]\rho_{jk}(t)\,.
\label{eq:EoM1}
\ee
Here
\be
L_{\rm coh}=\frac{2\sqrt{2}}{|\vec{v}_j-\vec{v}_k|}\sigma_x
\label{eq:Lcoh}
\ee
is the coherence length, i.e.\ the distance over which the WPs of different
neutrino mass eigenstates separate by a distance of the order of their spatial
length $\sigma_x$. Eq.~(\ref{eq:EoM1}) resembles the standard evolution
equation for the fixed-momentum components of the density matrix, except that
the usual energies $E_j$ are replaced by the effective ones $\tilde{E}_j$ and
that an extra term (the last term in the square brackets) appears. It should be
stressed, however, that we are now considering the density matrix of the
neutrino as a whole, which includes integration over all its momentum modes. It
is easy to see that the extra term in (\ref{eq:EoM1}) leads to damping with
time of the off-diagonal elements of $\rho_{jk}(t)$, which is a consequence of
the spatial separation of the WPs of different mass eigenstates.  Note also
that eq.~(\ref{eq:expand2}) implies
\be
\tilde{E}_j-\tilde{E}_k\simeq \frac{\Delta m^2}{2E}\,,
\label{eq:DeltaE}
\ee
which is just the quantity relevant for neutrino oscillations.

It is also interesting to study the EoM for the un-averaged density matrix
$\rho_{jk}(t,\vec{x})$. From eq.~(\ref{eq:gauss3}) we obtain
\be
\frac{\partial}{\partial t}\rho_{jk}(t, \vec{x})=
\Big[-i(E_j-E_k)+\frac{1}{\sigma_x^2}(\vec{x}-\vec{v}_g t)\vec{v}_g
-\frac{(\vec{v}_j-\vec{v}_k)^2}{4\sigma_x^2}t
\Big]\rho_{jk}(t, \vec{x})\,,
\vspace*{1.0mm}
\label{eq:EoM2a}
\ee
\be
\vec{\nabla}\rho_{jk}(t, \vec{x})=
\Big[i(\vec{p}_j-\vec{p}_k)-\frac{1}{\sigma_x^2}(\vec{x}-\vec{v}_g t)\Big]
\rho_{jk}(t, \vec{x})\,.
\qquad\qquad
\qquad\qquad~~\,
\label{eq:EoM2b}
\ee
Combining these two equations, we find
\be
\Big(\frac{\partial}{\partial t}+\vec{v}_g
\vec{\nabla}\Big)
\rho_{jk}(t, \vec{x}) = \Big[-i(\tilde{E}_j-\tilde{E}_k)-\frac{2t}
{L_{\rm coh}^2}\Big]\rho_{jk}(t, \vec{x})\,,
\label{eq:EoM3}
\ee
where we have used eq.~(\ref{eq:not1}). Recall now that the total time 
derivative, when applied to a classical function of time, coordinate and 
momentum, can be written as the Liouville operator
\be
\frac{d}{dt}\,=\,
\frac{\partial}{\partial t}+\dot{\vec{x}}\cdot
\vec{\nabla}
+\dot{\vec{p}}\cdot
\frac{\partial}{\partial \vec{p}}\,.
\label{eq:Liouv}
\ee
If we replace here $\dot{\vec{x}}$ with the average group velocity $\vec{v}_g$
and take into account that for free neutrinos (as well as for neutrinos
experiencing only forward scattering) the neutrino momentum is conserved, we
find that this operator formally coincides with the one acting on $\rho_{jk}(t,
\vec{x})$ on the left hand side of (\ref{eq:EoM3}).  Modulo the identification
of this operator with the total time derivative, eq.~(\ref{eq:EoM3}) would
essentially coincide with eq.~(\ref{eq:EoM1}), i.e.\ the averaged and
un-averaged density matrices would satisfy the same EoM.   This fact is rather
curious: strictly speaking, the Liouville operator is only applicable for
pointlike particles moving along classical trajectories, so that
$\vec{x}=\vec{x}(t)$.  But from our considerations it follows that it still
applies to the case of particles represented by finite-size WPs, provided that
their motion is described in terms of the average group velocity.

It is easy to show that our evolution equation (\ref{eq:EoM1})
actually follows from the standard momentum-space
EoM of the fixed-momentum components of the neutrino density matrix. (We shall
demonstrate this in Section~\ref{sec:adiab} in the more general case of neutrino
propagation in medium in the adiabatic regime.) The advantage of
eq.~(\ref{eq:EoM1}) for studying propagation decoherence is that it allows one
to immediately see the effects of the spatial separation of the WPs of the
different mass eigenstates, without integration over different neutrino
momentum modes.

At this point, some comments are in order.
\begin{enumerate}
  \item
  We have derived the EoM (\ref{eq:EoM1}) for $\rho_{jk}$ simply by
  differentiating the already known solution.  Of course, if the solution is
  known, we do not need any EoM.  Our point here was just to demonstrate that
  the fact that the off-diagonal elements of $\rho_{jk}$ decrease with time is
  reflected in the appearance of an extra term in the EoM, provided that we
  consider as our object of interest the density matrix of an individual
  neutrino as a whole, rather than the density matrices of its specific
  momentum modes.  The density matrix for a neutrino ensemble can then be
  obtained by summing over all neutrinos in the ensemble.

\item
The form of the damping term in eq.~(\ref{eq:EoM1}) depends on the assumed
shape of the neutrino WPs.  For example, for exponential rather than Gaussian
shape of the coordinate-space neutrino WPs,%
\footnote{Neutrinos with such WPs
are produced, e.g., in decays of free or quasi-free parent particles
\cite{HS,AHS,MSm}.} 
the extra term  would have the form $-1/L_{\rm coh}$ rather
than $-2t/L_{\rm coh}^2$, as in eq.~(\ref{eq:EoM1}) \cite{AKL3}.  This leads to
a dependence of the solutions of the neutrino evolution equation on the shape 
of the WPs at short and intermediate times; the solutions at asymptotically 
large times, however, exhibit vanishing off-diagonal elements of $\rho_{jk}$ 
whenever the damping term is present in the EoM, irrespectively of the exact
form of this term.

  \item
  The suppression with time of the off-diagonal elements of the neutrino density
  matrix in the mass eigenstate basis $\rho_{jk}$ is the direct consequence of
  the fact that its elements with $j \ne k$ contain products of the wave
  functions of the neutrino mass eigenstates propagating with different group
  velocities (see eq.~(\ref{eq:rho2a})).  Therefore, their overlap, which is
  maximal at neutrino production, decreases with time.  This suppression, which
  can be immediately seen from the expression for the coordinate-integrated
  density matrix $\rho(t)$ in eq.~(\ref{eq:rho4}), is actually present in the
  un-integrated matrix (\ref{eq:gauss3}) as well.  This can be readily seen if we
  rewrite in the 
exponent of (\ref{eq:gauss3})
  \be
  -\frac{1}{4\sigma_x^2}\big[
  (\vec{x}-\vec{v}_j t)^2+(\vec{x}-\vec{v}_k t)^2\big]=
  -\frac{(\vec{x}-\vec{v}_g t)^2}{2\sigma_x^2}-\frac{(\vec{v}_j-\vec{v}_k)^2 t^2}
  {8\sigma_x^2}\,.
  \label{eq:new}
  \ee

  \item
  Physically, the reason for the suppression of $\rho_{jk}$ with $j\ne k$
  with time is the separation of the WPs corresponding to different
  propagation-eigenstate components of each neutrino.  One might consequently
  expect this phenomenon to occur also in more general situations -- for
  neutrinos propagating in ordinary matter or in dense neutrino gases.  As we
  shall see in Sections~\ref{sec:adiab} and~\ref{sec:model2}, this is indeed
  the case in the adiabatic neutrino propagation regime, but in general is
  not true when adiabaticity is violated.
\end{enumerate}

\subsection{\label{sec:medium}Oscillations in normal matter
            and neutrino backgrounds}

We now turn to neutrino oscillations in normal matter and dense neutrino
environments in the WP picture.  Let us first consider evolution of neutrinos
of a given momentum $\vec{p}$ in the propagation eigenstate basis.%
\footnote{Oscillations in dense neutrino gases have been considered from the
viewpoint of the propagation eigenstate basis in \cite{cristina1,cristina2},
but the issue of WPs and their separation has not been addressed there.}
As was discussed in Section \ref{sec:stvec}, the propagation eigenstates
$\nu_j$ are the instantaneous eigenstates of the Hamiltonian of the neutrino
system ${\cal H}_{\vec{p}}(t)$, i.e.\ the states that diagonalize ${\cal
H}_{\vec{p}}(t)$ at the time $t$. We shall be using the matrix 
representation~(\ref{eq:H1}) of the neutrino Hamiltonian, and the column 
vector $\Psi$, the elements of which are the probability amplitudes of finding 
the corresponding flavor eigenstate neutrino at time $t$, to describe the 
neutrino state. The element $\Psi_\beta$ of the vector $\Psi$ is related to 
the state vector in the Hilbert space $|\nu(t, \vec{p})\rangle$ by
\be
\Psi_\beta(t, \vec{p})=\langle \nu_\beta|\nu(t, \vec{p})\rangle\,,
\label{eq:Psi}
\ee
and similarly for $|\nu^{p}\rangle$ and $\Psi^p$ describing the propagation 
eigenstates. Taking into account that 
the two bases are related by the leptonic mixing matrix $U(t)$, we find
from EoM~(\ref{eq:Schr1}) the neutrino evolution equation in the propagation
eigenstate basis: 
\be
i\frac{d}{dt} \Psi^p = \big[{\cal H}^d(t)-iU^\dag(t)\dot{U}(t)\big]
\Psi^p \equiv \tilde{{\cal H}}(t) \Psi^p\,.
\label{eq:Schr2}
\ee
Here the diagonal matrix ${\cal H}^d(t)$ was defined in eqs.~(\ref{eq:H2m})
and (\ref{eq:eigen}), and we suppressed the dependence of the Hamiltonians on
the neutrino momentum to simplify the notation.

Note that, unlike ${\cal H}^d(t)$, the effective Hamiltonian $\tilde{{\cal
H}}(t)$ that governs the evolution of the propagation eigenstates is in general
not diagonal because $U^\dag(t)\dot{U}(t)$ is not. This is related to the fact
that the Hamiltonian ${\cal H}(t)$ cannot be diagonalized by the same unitary
transformation at all times. If, however, the properties of the medium vary
slowly enough along the neutrino path, the neutrino system has enough time to
`adjust' itself to the changes of these properties.  In this case the term
$U^\dag(t)\dot{U}(t)$ in (\ref{eq:Schr2}) can be neglected, i.e.\ the system
evolves adiabatically.  The effective Hamiltonian $\tilde{\cal H}(t)$ then
essentially coincides with ${\cal H}^d(t)$, i.e.\ is diagonal. This means that
transitions between different propagation eigenstates are strongly suppressed,
and they evolve independently of each other.

\subsubsection{\label{sec:adiab}Adiabatic evolution}

Let us first consider neutrino flavour evolution in normal matter and
neutrino backgrounds in the adiabatic regime.
Let a neutrino be produced at a time $t_0$ and coordinate $\vec{x}_0$
in the flavour state $\alpha$ ($\alpha = e, \mu$ or $\tau$). According to
eq.~(\ref{eq:expan2}), its state vector can be written as a linear
combination of the state vectors of the propagation eigenstates.
Consider now the time evolution of the produced neutrino state.
In the adiabatic regime
the propagation eigenstates evolve independently, and their
time evolution is very simple -- they just acquire phase factors:
\be
t_0\to t:\qquad
|\nu_j^{(0)}(t_0,\vec{p})\rangle \;\,\to\;\,
e^{-i\int_{t_0}^t E_j(t',\,\vec{p})dt'}|\nu_j^{(0)}(t,\vec{p})\rangle.
\label{eq:evol1}
\ee
This equation should be understood as follows. In general, the
state to which $|\nu_j^{(0)}(t_0,\vec{p})\rangle$ evolves as time advances from
$t_0$ to $t$ can be expanded in the basis of the eigenstates
$|\nu_k^{(0)}(t,\vec{p})\rangle$ of the Hamiltonian ${\cal H}(t)$. In the
adiabatic regime,
this expansion contains only one term (with $k=j$), the expansion coefficient
being the exponential factor in (\ref{eq:evol1}).

The evolved neutrino state
can now be obtained by applying eq.~(\ref{eq:evol1}) to
(\ref{eq:expan3}):
\be
|\nu(t,\vec{p})\rangle=
\sum_je^{-i\int_{t_0}^t E_j(t',\vec{p})dt'}f_{\vec{p}_j}(\vec{p})
U^*_{\alpha j}(t_0)\,|\nu_j^{(0)}(t,\vec{p})\rangle.
\label{eq:wp2m}
\ee
Substituting this into (\ref{eq:wp1m}), we obtain
\be
|\nu(t,\vec{x})\rangle=\sum_j\int\frac{d^3 p}{(2\pi)^3}\,
e^{-i\int_{t_0}^t E_j(t',\vec{p})dt'+i\vec{p}\vec{x}'}
f_{\vec{p}_j}(\vec{p})U^*_{\alpha j}(t_0)\,|\nu_j^{(0)}
(t,\vec{p})\rangle\,,
\label{eq:wp3m}
\ee
where
\be
\vec{x} '\equiv\vec{x}-\vec{x}_0\,.
\label{eq:xprime}
\ee
For the 1-particle neutrino density matrix in the propagation eigenstate
basis we then obtain
\begin{align}
\rho_{jk}(t,\vec{x})=
\int\frac{d^3p}{(2\pi)^3}\frac{d^3p'}{(2\pi)^3}
e^{-i\int_{t_0}^t[E_j(t',\vec{p})-E_k(t',\vec{p}\,')]dt'
+i(\vec{p}-\vec{p}\,')\vec{x}'}
f_{\vec{p}_j}(\vec{p})f^*_{\vec{p}_k}(\vec{p}\,')
U_{\alpha j}^*(t_0)U_{\alpha k}^{}(t_0)\,.
\label{eq:rho1a}
\end{align}
Notice that the quantities $U_{\alpha j}(t_0)$ and $U_{\alpha k}(t_0)$ here
implicitly depend on $\vec{p}$ and $\vec{p}\,'$, respectively.

It is interesting to note that in the special case when the initially
produced neutrino flavour state coincides at
$t=t_0$ with one of the instantaneous propagation eigenstates, i.e.\
$U_{\alpha j}(t_0)=\delta_{\alpha j}$, the off-diagonal elements of the
propagation-basis density matrix $\rho_{jk}(t,\vec{x})$ vanish at all times.
This, however, is only true in the adiabatic approximation.

Let us expand $E_j(t,\vec{p})$ and $E_k(t,\vec{p}\,')$ in the integrand of
eq.~(\ref{eq:rho1a}) near the peaks of the corresponding momentum
distribution functions and keep only the first two terms in the expansion.%
\footnote{\label{foot:spread}
The higher-order terms are responsible for the spreading of the wave packets, 
which we neglect here.} 
This yields
\be
E_j(t,\vec{p})-E_k(t,\vec{p}\,')\simeq
E_j(t)-E_k(t)+\vec{v}_{j}(t)(\vec{p}-\vec{p}_j)-
\vec{v}_{k}(t)(\vec{p}\,'-\vec{p}_k)\,,
\label{eq:expand1m}
\ee
where
\be
E_j(t)\equiv E_j(t, \vec{p}_j)\,,\qquad
\vec{v}_j(t)\equiv \frac{\partial E_j(t,\vec{p})}{\partial \vec{p}}
\big{|}_{\vec{p}=\vec{p}_j}\,,
\label{eq:Ev}
\ee
and similarly for $E_k(t)$ and $\vec{v}_k(t)$. Note that $\vec{v}_j(t)$ is
the group velocity of the $j$th neutrino propagation eigenstate. In general,
it is time-dependent in medium because so is the effective Hamiltonian
${\cal H}_{\vec{p}}(t)$. In the following,
it will be convenient for us to introduce the mean group velocities
$\langle\vec{v}_j(t)\rangle$ according to
\be
\langle\vec{v}_j(t)\rangle\,\equiv\, \frac{1}{t-t_0}
\int_{t_0}^t \vec{v}_j(t')dt'\,.
\label{eq:mean}
\ee
Substituting the expansion (\ref{eq:expand1m}) into eq.~(\ref{eq:rho1a}),
we find
\be
\rho_{jk}(t,\vec{x})\simeq
U_{\alpha j}^*(t_0)U_{\alpha k}^{}(t_0)
e^{-i\int_{t_0}^t[E_j(t')-E_k(t')]dt'
+i(\vec{p}_j-\vec{p}_k)\vec{x}'}
\qquad\qquad\qquad
\qquad\qquad\qquad
\qquad\qquad
\nonumber
\vspace*{-0.8mm}
\ee
\be
\qquad
\times \int\frac{d^3q}{(2\pi)^3}\frac{d^3q'}{(2\pi)^3}
e^{i\vec{q}\big[\vec{x}'
-\langle\vec{v}_j(t)\rangle(t-t_0)\big]
-i\vec{q}\,'\big[\vec{x}'-\langle\vec{v}_k(t)\rangle(t-t_0)
\big]}
f_{\vec{p}_j}(\vec{p}_j+\vec{q})f^*_{\vec{p}_k}(\vec{p}_k+\vec{q}\,').
\label{eq:rho3a}
\ee
Here the integration is performed over the shifted momenta
$\vec{q}=\vec{p}-\vec{p}_j$ and $\vec{q}\,'=\vec{p}\,'-\vec{p}_k$. We have
also taken into account that the leptonic mixing matrix in medium $U(t)$
varies little over momentum intervals comparable to the width $\sigma_p$ of
the momentum distribution function $f_{\vec{p}_j}(\vec{p})$;
this allowed us to replace $U_{\alpha j}^*(t_0)$ and $U_{\alpha k}^{}(t_0)$
by their values taken, respectively, at $\vec{p}=\vec{p}_j$ and
$\vec{p}\,'=\vec{p}_k$ and pull them out of the momentum integrals.

The integrals over momenta in (\ref{eq:rho3a}) factorize, and we obtain
\begin{align}
\rho_{jk}(t,\vec{x})\simeq U_{\alpha j}^*(t_0)U_{\alpha k}^{}(t_0)
e^{-i\int_{t_0}^t[E_j(t')-E_k(t')]dt'
+i(\vec{p}_j-\vec{p}_k)\vec{x}'}
\qquad\qquad
\nonumber \\[3mm]
\times g_j\big[\vec{x}'-\langle\vec{v}_j(t)\rangle(t-t_0)\big] \,
g_k^*\big[\vec{x}'-\langle\vec{v}_k(t)\rangle(t-t_0)\big],
\label{eq:rho4m}
\end{align}
where
\be
g_j\left[\vec{x}'-\langle\vec{v}_j(t)\rangle(t-t_0)\right]\equiv
\int\frac{d^3q}{(2\pi)^3}
e^{i\vec{q}\big[\vec{x}'-\langle\vec{v}_j(t)\rangle(t-t_0)\big]}
f_{\vec{p}_j}(\vec{p}_j+\vec{q})
\label{eq:gj}
\ee
is the envelope function of the coordinate-space WP of $\nu_j$. In particular,
for Gaussian momentum-space neutrino WPs (\ref{eq:gauss1m}),
the envelope functions are also Gaussian:
\be
g_j\left[\vec{x}'-\langle\vec{v}_j(t)\rangle(t-t_0)\right]=
(2\pi\sigma_x^2)^{-3/4}\,
\exp\Big[-\frac{\left[\vec{x}'-\langle\vec{v}_j(t)\rangle(t-t_0)
\right]^2}{4\sigma_x^2}\Big]\,,
\qquad \sigma_x\equiv\frac{1}{2\sigma_p}\,.
\label{eq:gauss2m}
\ee

Let us now go to the averaged neutrino density matrix $\rho(t)$ by
integrating $\rho(t,\vec{x})$ over spatial regions of linear size $\bar{r}
\gg \sigma_x$ around $\vec{x}$, i.e.\ by performing coarse graining.
As was discussed in Section~\ref{sec:vac},
since the coordinate-space WPs quickly decrease at distances from their centers
that are large compared with $\sigma_x$, one can formally extend the
integration over $\vec{x}$ to an infinite spatial volume.
Eq.~(\ref{eq:rho4m}) then yields
\be
\rho_{jk}(t)=
U_{\alpha j}^*(t_0)U_{\alpha k}^{}(t_0)
e^{-i\int_{t_0}^t[E_j(t')-E_k(t')]dt'} G_{jk}(t)\,,
\label{eq:rho6}
\ee
where $\rho_{jk}(t)$ was defined in eq.~(\ref{eq:rho3}) and
\be
G_{jk}(t)\,\equiv\,\int d^3x'\,
e^{i(\vec{p}_j-\vec{p}_k)\vec{x}'}
g_j\left[\vec{x}'-\langle\vec{v}_j(t)\rangle(t-t_0)\right]
g_k^*\left[\vec{x}'-\langle\vec{v}_k(t)\rangle(t-t_0)\right].
\label{eq:G1}
\ee
Note that, while
$\rho_{jk}(t, \vec{x})$ corresponds to a pure state (i.e.\ its $j$- and
$k$-dependence factorizes), this is not true for the spatially averaged
quantity $\rho_{jk}(t)$.
It is important to emphasize that, strictly speaking, the `coarse-grained'
quantity (\ref{eq:rho6}) depends solely on time only in the case of uniform
ordinary matter and neutrino backgrounds; otherwise, it will
also depend on the coordinate of the point around which the coarse graining
with $\bar{r}\gg \sigma_x$ is performed. In what follows,
we will concentrate on the case of uniform media, though, with minimal
modifications, some of our results will apply to the case of non-uniform
backgrounds as well.

Let us introduce the notation
\be
\langle\vec{v}_g(t)\rangle 
\equiv \frac{\langle\vec{v}_j(t)\rangle+\langle\vec{v}_k(t)
\rangle}{2}\,,\quad\Delta\vec{v}(t)\equiv \langle\vec{v}_j(t)\rangle-
\langle\vec{v}_k(t)
\rangle\,,\quad
\vec{x}_1\equiv \vec{x}' -
\langle\vec{v}_g(t)\rangle (t-t_0)\,.
\label{eq:not1m}
\ee
Then we can rewrite eq.~(\ref{eq:G1}) as
\be
G_{jk}(t)\,=\,
e^{i(\vec{p}_j-\vec{p}_k)
\langle\vec{v}_g(t)\rangle (t-t_0)}\vspace*{3.5mm}F_{jk}(t)\,,
\vspace*{-2.5mm}
\qquad\qquad\qquad\qquad\qquad\qquad\qquad\qquad\quad
\label{eq:G2}
\ee
\be
F_{jk}(t)\equiv\int d^3x_1\,
e^{i(\vec{p}_j-\vec{p}_k)\vec{x}_1}
g_j\Big[\vec{x}_1-\frac{\Delta\vec{v}(t)}{2}(t-t_0)\Big]
g_k^*\Big[\vec{x}_1+\frac{\Delta\vec{v}(t)}{2}(t-t_0)\Big]\,,
\label{eq:F1}
\ee
and $\rho_{jk}(t)$ takes the form
\be
\rho_{jk}(t)=
U_{\alpha j}^*(t_0)U_{\alpha k}^{}(t_0)
e^{-i\int_{t_0}^t[\tilde{E}_j(t')-\tilde{E}_k(t')]dt'}F_{jk}(t)\,.
\label{eq:rho7m}
\ee
Here, in analogy with eq.~(\ref{eq:not1}), we have defined
\be
\tilde{E}_j(t) \equiv E_j(t)-\vec{v}_g(t)\vec{p}_j
\qquad \text{with}\qquad \vec{v}_g\equiv\frac{\vec{v}_j(t)+\vec{v}_k(t)}{2}\,.
\label{eq:not2}
\ee
The envelope function $g_j(\vec{z})$ is peaked at or close to the zero of
its argument and decreases rapidly when the argument becomes larger than the
spatial width of the WP $\sigma_x$. The splitting of the arguments of
$g_j$ and $g_k^*$ in the integrand of (\ref{eq:F1}) is $\Delta v(t)(t-t_0)$;
when this splitting exceeds $\sigma_x$ (i.e., $t-t_0$ exceeds the coherence
length), the overlap of $g_j$ and $g_k^*$
gets suppressed, leading to the suppression of the off-diagonal elements
of $G_{jk}(t)$.

Let us now find the EoM satisfied by $\rho_{jk}(t)$. From
eq.~(\ref{eq:rho7m}) we obtain
\be
\dot{\rho}_{jk}(t) =
-i\big[\tilde{E}_j(t)-\tilde{E}_k(t)\big]
\rho_{jk}(t)
+\frac{\dot{F}_{jk}}{F_{jk}}\rho_{jk}(t)\,,
\label{eq:EoM1m}
\ee
where as usual dots denote time derivatives.
The last term in eq.~(\ref{eq:EoM1m})
describes the suppression with time of the off-diagonal elements of the
neutrino density matrix in the propagation eigenstate basis. In particular,
for Gaussian WPs (\ref{eq:gauss2m}) we find
\be
F_{jk}(t) =
\exp\Big[-\frac{(\vec{p}_j-\vec{p}_k)^2}{8\sigma_p^2}\Big]
\exp\Big[-\frac{[\Delta\vec{v}(t)]^2 (t-t_0)^2}
{8\sigma_x^2}\Big].
\label{eq:F2}
\ee
Eq.~(\ref{eq:EoM1m}) then yields
\be
\dot{\rho}_{jk}(t) = \Big[-i(\tilde{E}_j-\tilde{E}_k)-\frac{2(t-t_0)}
{L^2_{\rm coh}(t)}\Big]\rho_{jk}(t)\,,
\label{eq:EoM2}
\ee
where the coherence length is given by
\be
L_{\rm coh}(t)=\frac{2\sqrt{2}\sigma_x}{\sqrt{[\vec{v}_j(t)-\vec{v}_k(t)]
[\langle\vec{v}_j(t)\rangle-\langle\vec{v}_k(t)\rangle]}}
=\frac{2\sqrt{2}\sigma_x}{\sqrt{\Delta\vec{v}(t)[\vec{v}_j(t)-\vec{v}_k(t)]}}\,.
\label{eq:Lcoh1}
\ee
Note that eq.~(\ref{eq:EoM2}) has a form similar to that in the case of
vacuum oscillations, eq.~(\ref{eq:EoM1}) (where $t_0$ was set equal to zero).
Interestingly, the expression for $L_{\rm coh}(t)$ in eq.~(\ref{eq:Lcoh1})
contains, along with the difference of the mean velocities
$\langle\vec{v}_{j}(t)\rangle-\langle\vec{v}_{k}(t)\rangle$,
the difference of the instantaneous time-dependent velocities
$\vec{v}_{k}(t)-\vec{v}_{k}(t)$. It was pointed out in \cite{MikhSm2} that in
the case of neutrino oscillations in ordinary matter this velocity difference
changes its sign close to the MSW resonance, which may in principle lead to
the complete or partial restoration of propagation coherence.
In Section~\ref{sec:coher} of the present paper it was shown
that a similar phenomenon takes place in the case of oscillations in
dense neutrino backgrounds. This, however, happens
in a very narrow region of parameter space, outside of which the time 
dependence of $\vec{v}_{j,k}(t)$ is very weak and the difference between
$\langle\vec{v}_{j}(t)\rangle-\langle\vec{v}_{k}(t)\rangle$ and
$\vec{v}_{k}(t)-\vec{v}_{k}(t)$ essentially disappears
(see Section~\ref{sec:coher}). In this case the
expression for $L_{\rm coh}(t)$ in eq.~(\ref{eq:Lcoh1}) simplifies to
\be
L_{\rm coh}(t)=\frac{2\sqrt{2}}{|\Delta\vec{v}(t)|}\sigma_x\,,
\label{eq:Lcoh2}
\ee
which is similar to eq.~(\ref{eq:Lcoh}).

For future discussion and to compare our results with those obtained in the
standard approach,%
\footnote{We recall that by the standard approach we mean the 
formalism in which individual neutrino momentum modes are evolved in time 
without reference to wave packets. In momentum space, introducing wave packets 
corresponds to coarse graining in the momentum variable.}
it will be convenient to express $\rho_{jk}(t)$ and the
time derivatives of $\rho_{jk}(t, \vec{x})$ and $\rho_{jk}(t)$ also directly
from eq.~(\ref{eq:rho1a}), taking the integral over the coordinates
in the expression for $\rho_{jk}(t)$ and $\dot{\rho}_{jk}(t)$ before the
momentum integrals. For $\rho_{jk}(t)$ we find
\be
\rho_{jk}(t)=
\int\frac{d^3p}{(2\pi)^3}
e^{-i\int_{t_0}^t[E_j(t',\vec{p})-E_k(t',\vec{p})]dt'}
f_{\vec{p}_j}(\vec{p})f^*_{\vec{p}_k}(\vec{p})
U_{\alpha j}^*(t_0)U_{\alpha k}^{}(t_0)\,.
\label{eq:rho8}
\ee
Differentiating (\ref{eq:rho1a}) and (\ref{eq:rho8}) with respect to time
yields, respectively,
\begin{align}
\dot{\rho}_{jk}(t, \vec{x})=
-i\int\frac{d^3p}{(2\pi)^3}\frac{d^3p'}{(2\pi)^3}
[E_j(t,\vec{p})-E_k(t,\vec{p}\,')]e^{-i\int_{t_0}^t
[E_j(t',\vec{p})-E_k(t',\vec{p}\,')]dt'
+i(\vec{p}-\vec{p}\,')\vec{x}}
\nonumber \\
\times f_{\vec{p}_j}(\vec{p})f^*_{\vec{p}_k}(\vec{p}\,')
U_{\alpha j}^*(t_0)U_{\alpha k}^{}(t_0),
\label{eq:EoM3m}
\end{align}
\be
\dot{\rho}_{jk}(t)=-i\int\frac{d^3p}{(2\pi)^3}
\big[E_j(\vec{p})-E_k(\vec{p})\big]
e^{-i\int_{t_0}^t[E_j(t',\vec{p})-E_k(t',\vec{p})]dt'}
f_{\vec{p}_j}(\vec{p})f^*_{\vec{p}_k}(\vec{p})
U_{\alpha j}^*(t_0)U_{\alpha k}^{}(t_0)\,.
\label{eq:EoM4m}
\ee
Note that $\rho_{jk}(t)$ can be obtained either as the integral of
$\rho_{jk}(t,\vec{x})$ over the coordinate or, equivalently, as the integral
of its Fourier transform $\rho_{jk}(t,\vec{p})$ over momentum; it is then
clear that the integrand in eq.~(\ref{eq:rho8}) is just
$\rho_{jk}(t,\vec{p})$.
The integrand of eq.~(\ref{eq:EoM4m}) therefore represents the
right hand side of the standard EoM
for the fixed-momentum components of the density matrix
\be
\dot{\rho}(t,\vec{p})=-i\big[{\cal H}_{\vec{p}}(t), \rho(t,\vec{p})\big]
\label{eq:SR}
\ee
written in the neutrino propagation-eigenstate basis.
Indeed, in the adiabatic regime the Hamiltonian $\tilde{\cal H}_{\vec{p}}(t)$ is
diagonal in the propagation eigenstate basis, and the commutator in
eq.~(\ref{eq:SR}) leads to the factor $\big[E_j(\vec{p})-E_k(\vec{p})\big]$
in the integrand of eq.~(\ref{eq:EoM4m}), whereas the rest of
the integrand is just $\rho_{jk}(t,\vec{p})$.
Although at first sight eq.~(\ref{eq:EoM4m}) does
not have a damping term leading to the suppression of the off-diagonal elements
of $\rho_{jk}(t)$, such a damping is  actually present there.
Indeed, neutrino propagation decoherence
can be described either in momentum space or in configuration
space. In momentum space it
stems from the dephasing of different
neutrino modes at late times and is related to the fast oscillations
of the integrand in the integral over the neutrino spectrum
in~(\ref{eq:rho8}). In configuration space decoherence
is related to the spatial separation of the WPs of different
propagation eigenstates after they
have traveled long enough distance. The momentum space and configuration space
descriptions are equivalent (see, e.g., \cite{parad}). In the case
we consider here, this follows from the fact that eq.~(\ref{eq:rho8}) is
equivalent to eq.~(\ref{eq:rho7m}), whose time derivative (\ref{eq:EoM1m})
contains the damping term $\propto \dot{F}_{jk}/F_{jk}$.

\subsubsection{\label{sec:nonadiab}Beyond the adiabatic approximation}

Consider now non-adiabatic neutrino flavour evolution. It is important to note
that,  unlike for neutrino oscillations in ordinary matter,
in dense neutrino environments adiabaticity violation may take place even in
the case of constant neutrino density. This comes about because in this
case the Hamiltonian of the system depends not only on the overall density of
the neutrino gas, but also on its flavour composition, which changes with time.

If neutrino evolution is non-adiabatic, the evolution of different propagation
eigenstates is not independent. This means that a propagation eigenstate
$\nu_i$ produced at a time $t_0$ will at a later time $t$ become a linear
superposition of different propagation eigenstates:
\be
t_0\to t:
\quad\quad
|\nu_i^{(0)}(t_0,\vec{p})\rangle
\;\,\to\;\,
\sum_j
S_{ji}(t,t_0;\vec{p})|\nu_j^{(0)}(t,\vec{p})\rangle\,.
\vspace*{-2mm}
\label{eq:evol2}
\ee
The matrix $S_{ji}(t,t_0;\vec{p})$ is the neutrino evolution matrix in the
propagation eigenstate basis, i.e.\ the evolution matrix that solves
eq.~(\ref{eq:Schr2}).
It is related to the corresponding evolution matrix in the flavour basis
$S^{\rm fl}(t,t_0;\vec{p})$ by
\be
S^{\rm fl}(t,t_0;\vec{p})=U(t)S(t,t_0;\vec{p})U^\dag(t_0)\,.
\label{eq:Sfl}
\ee
Applying eq.~(\ref{eq:evol2}) to (\ref{eq:expan3}), we
obtain the evolved neutrino state vector in momentum space:
\be
|\nu(t,\vec{p})\rangle=
\sum_{i,j}
S_{ji}(t,t_0;\vec{p})f_{\vec{p}_i}(\vec{p})
U^*_{\alpha i}(t_0)\,|\nu_j^{(0)}(t,\vec{p})\rangle.
\vspace*{-2mm}
\label{eq:wp4}
\ee
Following then the same steps as in Section \ref{sec:adiab}, for the
coordinate-averaged 1-particle neutrino density matrix in the
propagation eigenstate basis we find
\be
\rho_{jk}(t)=\sum_{i,l}
\int\frac{d^3p}{(2\pi)^3}
\,S_{ji}(t,t_0;\vec{p})S_{kl}^*(t,t_0;\vec{p})
f_{\vec{p}_i}(\vec{p})f^*_{\vec{p}_l}(\vec{p})
U_{\alpha i}^*(t_0)U_{\alpha l}^{}(t_0)\,.
\label{eq:rho10}
\ee
In the adiabatic limit, $S_{ji}(t,t_0;\vec{p})=e^{-i\int_{t_0}^t
E_j(t',\vec{p})dt'}\delta_{ji}$, and we recover  eq.~(\ref{eq:rho8}).

It is important that, unlike in the adiabatic case, the off-diagonal elements
of the density matrix (\ref{eq:rho10}) do not in general vanish when the
initially produced neutrino state coincides with one of the instantaneous
propagation eigenstates, i.e.\ when $U_{\alpha i}(t_0)=\delta_{\alpha i}$.
Indeed, in this case
\be
\rho_{jk}(t)=
\int\frac{d^3p}{(2\pi)^3}
\,S_{j\alpha}(t,t_0;\vec{p})S_{k\alpha}^*(t,t_0;\vec{p})
|f_{\vec{p}_\alpha}(\vec{p})|^2\,,
\label{eq:rho10a}
\ee
and its $j\ne k$ elements are obviously non-zero in the non-adiabatic
case.

Let us now consider
neutrino evolution in the non-adiabatic regime in more detail.
An important point is that in the presence of neutrino self-interactions the
Hamiltonian ${\cal H}(t)$ is in general complex (though is still, of course,
Hermitian) even in the absence of fundamental
CP violation. Indeed, as follows from eq.~(\ref{eq:H1}), ${\cal H}(t)$ depends
on the neutrino density matrix $\rho$.
Since the off-diagonal elements of $\rho$ are in general complex, so
is ${\cal H}(t)$.%
This, in particular, means that ${\cal H}(t)$ is diagonalized
by a complex unitary transformation rather than by a real orthogonal one.
This applies even to the 2-flavour case, in which in the absence of neutrino
self-interactions the mixing matrix $U$ contains no physical complex
phases. Indeed, since the phase
stemming from neutrino-neutrino
interactions is time-dependent, it cannot be rotated away by a simple
time-independent redefinition of one of the neutrino fields; therefore,
it affects their time-dependent phase difference and hence the probabilities
of the flavour transitions.

For simplicity, we will confine our consideration to the 2-flavour case.
The unitary matrix $U(t)$ diagonalizing ${\cal H}(t)$ can then be taken in the
form
\be
U(t)=
\left(\begin{array}{cc}
1 & 0  \\
0 & e^{-i\phi(t)}  \end{array} \right)
\left(\begin{array}{cc}
c & s  \\
-s & c  \end{array} \right),
\label{eq:U1}
\ee
where $c\equiv\cos\theta(t)$ and $s\equiv\sin\theta(t)$. The angle $\theta(t)$
and the phase $\phi(t)$ can be expressed through the matrix elements of
${\cal H}(t)$:
\be
\tan 2\theta = \frac{2|{\cal H}_{12}|}{{\cal H}_{22}-{\cal H}_{11}}\,,\qquad
\phi=\arg({\cal H}_{12})\,.
\label{eq:angles1}
\ee
Next, we recall that for the purposes of studying neutrino flavour transitions
one can always choose ${\cal H}(t)$ to be traceless by using the transformation
${\cal H}(t)\to {\cal H}(t)-\tfrac{1}{2} tr[{\cal H}(t)]\!\cdot\!\mathbbm{1}$.
Eq.~(\ref{eq:H2m}) then \vspace*{1mm}yields
\be
{\cal H}(t)=\frac{\Delta}{2}
\left(\begin{array}{cc}
-c_2~~~ & s_2 e^{i\phi}  \\
s_2 e^{-i\phi}~ & c_2  \end{array} \right),
\label{eq:H1mm}
\ee
\noindent
where
\be
\Delta\equiv E_2(t)-E_1(t)\,,\qquad c_2\equiv\cos 2\theta(t)\,,\qquad
s_2\equiv\sin 2\theta(t)\,.
\label{eq:notat1}
\ee

Let us now find the effective Hamiltonian in the propagation eigenstate basis
$\tilde{{\cal H}}(t)$.
From the definition of $\tilde{{\cal H}}(t)$ in eq.~(\ref{eq:Schr2})
and eq.~(\ref{eq:U1}) we obtain
\be
\tilde{{\cal H}}(t)=
\left(\begin{array}{cc}
-\frac{\Delta}{2}-s^2\dot{\phi} & c s \dot{\phi}-i\dot{\theta}  \\
c s \dot{\phi}+i\dot{\theta} & \frac{\Delta}{2}-c^2\dot{\phi}
\end{array} \right).
\label{eq:H3}
\ee
As usual, the probability of transitions between different propagation
eigenstates (i.e.\ the degree of adiabaticity violation)
can be characterized by their mixing, i.e.\ by
the modulus of the ratio of twice the off-diagonal element of
$\tilde{{\cal H}}(t)$ to the difference of its diagonal elements.
The non-adiabaticity parameter is thus
\be
\lambda = \frac{\sqrt{s_2^2\dot{\phi}^2+4\dot{\theta}^2}}
{|\Delta-c_2\dot{\phi}|}.
\label{eq:lambda}
\ee
For $\lambda \ll  1$, transitions between the propagation eigenstates
are strongly suppressed and they evolve practically independently,
whereas for $\lambda \gtrsim 1$ strong transitions between them
are possible.

The appearance of $\dot{\phi}$ in the
expression for $\lambda$ is easy to interpret in the framework of the flavour
spin description of neutrino flavour transitions.
As discussed in Section \ref{sec:size} and in the Appendix, for
each $\omega$-mode neutrino flavour evolution is described by the precession
of the flavour spin vector $\vec{P}_\omega$ around the (in general
time-dependent) Hamiltonian vector $\vec{H}_\omega$. The angle between 
$\vec{H}_\omega$ and the 3rd axis in flavour space (the polar angle) is given by
2$\theta(t)$, whereas $\phi(t)$ is the azimuthal angle characterizing the
direction of the projection of $\vec{H}_\omega$ on the 1--2  
plane (see the Appendix). Adiabaticity means that the
flavour spin vectors $\vec{P}_\omega$ are fast
precessing around their corresponding $\vec{H}_\omega$,
whereas the direction of each $\vec{H}_\omega$ changes relatively slowly in
flavour space. In that case the
flavour spins `track' the motion of the $\vec{H}_\omega$. For neutrino
oscillations in ordinary matter the
effective Hamiltonian ${\cal H}_\omega(t)$ is real, so that $\phi(t)\equiv
0$. In that case the vector $\vec{H}_\omega(t)$ always lies in the 1--3 plane
in flavour space, and its direction is fully characterized by the angle
$\theta(t)$. With changing matter density the direction of $\vec{H}_\omega$
changes in the 1--3 plane with speed $2\dot{\theta}$, and
the adiabaticity is good when this speed is small compared to the angular
velocity $\Delta$ of precession of the flavour spin vector $\vec{P}_\omega$
around $\vec{H}_\omega$. As a result, the (non)adiabaticity parameter is given
by the standard expression $\lambda = 2|\dot{\theta}|/|\Delta|$.

In the case of oscillations in neutrino backgrounds, each vector
$\vec{H}_\omega$ performs a 3-dimensional motion, and the speed of the
change of its direction depends both on $\dot{\theta}$ and on the angular
velocity $\dot{\phi}$ of the rotation of the projection of $\vec{H}_\omega$
onto the 1--2 plane. Therefore, knowing just $\dot{\theta}$ would not be 
sufficient to characterize the degree of adiabaticity violation during 
neutrino evolution. For example, for $\vec{H}_\omega$ precessing around the 
3rd axis, the angle $\theta$ remains constant, while $\phi$ changes.
Note that such a situation can be approximately realized
for synchronized oscillations in dense uniform and isotropic neutrino gas.
As shown in the Appendix, in this case $\dot{\theta}$ oscillates around zero
or small average, taking both positive and negative values, whereas
$\dot{\phi}$ oscillates around a non-zero value, remaining for most of the
parameter space sign-definite
(see eqs.~(\ref{eq:thetadot}) and~(\ref{eq:phidot})).
Therefore, the situation when $\theta\simeq \text{const.}$ while $\phi$
changes with time
obtains for $\dot\theta$ and $\dot\phi$ averaged over their fast oscillations.

The evolution matrix of propagation eigenstates $S_{ji}(t,t_0;\vec{p})$
cannot be found in a model independent way. To find out if damping due to WP
separation is operative in the non-adiabatic regime, we consider two
illustrative models of adiabaticity violation.

\subsubsection{\label{sec:model1}A model with short-time non-adiabaticity}

In this section we adopt the following simple model of adiabaticity violation:
a spatially uniform and isotropic 2-flavour neutrino
system in which neutrinos are produced in a flavour $\alpha$ at a time $t_0$,
propagate adiabatically until $t_1-\varepsilon$, experience extremely
non-adiabatic evolution during a short time interval
$[t_1-\varepsilon,\,t_1+\varepsilon]$, and then evolve again adiabatically
until time $t$.%
\footnote{Such a situation may be realized e.g.\ in supernovae when neutrinos
cross a propagating shock wave: the adiabaticity is then expected to be
good everywhere except inside the shock wave front. We thank Alexei Smirnov
for this observation.}
As follows from eq.~(\ref{eq:lambda}), extreme non-adiabaticity means that
during the interval $[t_1-\varepsilon,\,t_1+\varepsilon]$ the condition
\be
|\Delta-c_2\dot{\phi}|\ll \sqrt{s_2^2\dot{\phi}^2+4\dot{\theta}^2}
\label{eq:nonadiab}
\ee
is satisfied. Barring possible fine-tunings, this requires, in particular,
that during this interval either $|\dot{\phi}|\ll |\dot{\theta}|$, or
$c_2\ll 1$ (i.e. $\theta(t)\simeq \pi/4=\text{const.}$ and so
$|\dot{\theta}|\varepsilon\ll 1$). In
the rest of this subsection
we will assume that the first of these two possibilities is
realized, i.e.\ $\dot{\phi}$ can be neglected; the second possibility can be
considered quite analogously.

The eigenvalues and eigenstates of the effective Hamiltonian
${\cal H}_{\vec{p}}(t)$ before and after the non-adiabatic transition may be
drastically different.
Let us denote
\be
\eta_{1,2}=\exp\Big[-i\int_{t_0}^{t_1-\varepsilon}E_{1,2}(t',\vec{p})dt'\Big],
\qquad
\tilde{\eta}_{1,2}=\exp\Big[-i\int_{t_1+\varepsilon}^{t}E_{1,2}
(t',\vec{p})dt'\Big],
\label{eq:eta}
\vspace*{1.0mm}
\ee
\be
\alpha=\theta(t_1+\varepsilon)-\theta(t_1-\varepsilon)\,,
\label{eq:alpha}
\ee
where $\theta(t_1-\varepsilon)$ and $\theta(t_1+\varepsilon)$ are the values
of the in-medium mixing angle immediately before and immediately after the
non-adiabatic transition.
The evolution matrix for the propagation eigenstates can be written as
\be
S(t,t_0,\vec{p})=
\left(\begin{array}{cc}\tilde{\eta}_1 & 0 \\
0 &\tilde{\eta}_2 \end{array} \right)
\left(\begin{array}{cc}
\cos\alpha & -\sin\alpha  \\
\sin\alpha &\cos\alpha  \end{array} \right)
\left(\begin{array}{cc}\eta_1 & 0 \\
0 & \eta_2 \end{array} \right).
\label{eq:S1}
\ee
Here the rightmost, central and leftmost matrices describe, respectively, the
adiabatic evolution in the interval $[t_0, t_1-\varepsilon]$, extremely
non-adiabatic evolution during $[t_1-\varepsilon, t_1+\varepsilon]$, and then
again the adiabatic evolution  in the interval \,$[t_1+\varepsilon, t]$.
Direct calculation gives
\be
S(t,t_0,\vec{p})\!=\!\!
\left(\begin{array}{cc}\!
\tilde{\eta}_1 \eta_1
\cos\alpha
\;& -\tilde{\eta}_1\eta_2 \sin\alpha
\! \vspace*{1.5mm} \\
\tilde{\eta}_2\eta_1
\sin\alpha \;&
\tilde{\eta}_2\eta_2 \cos\alpha \!
\end{array} \right),
\label{eq:S2}
\ee
Note that due to the smallness of $\varepsilon$ one can actually
set $\varepsilon\to 0$ in the integration limits in (\ref{eq:eta}), so that
\be
S(t,t_0,\vec{p})\!=\!\!
\left(\begin{array}{cc}
\!
e^{-i\int_{t_0}^t E_1(t',\vec{p})dt'}\cos\alpha \;&
{}-e^{-i[\int_{t_0}^{t_1} E_2(t',\vec{p})dt'+
\int_{t_1}^{t} E_1(t',\vec{p})dt']}
\sin\alpha \! \vspace*{1.5mm} \\
\!e^{-i[\int_{t_0}^{t_1} E_1(t',\vec{p})dt'+
\int_{t_1}^{t} E_2(t',\vec{p})dt']}
\sin\alpha \;&
e^{-i\int_{t_0}^t E_2(t',\vec{p})dt'}\cos\alpha \!
\end{array} \right)\!.
\label{eq:S2a}
\ee

Let us now concentrate on the off-diagonal elements of $\rho_{jk}(t)$.
We will first consider the question of whether they
exist and are damped with time when the initially produced neutrino state
coincides with one of the instantaneous propagation eigenstates or when
different neutrinos are produced at different times and so the averaging over
$t_0$ has to be performed when going from the 1-particle neutrino density
matrix to the complete one.
Recall that in the adiabatic regime in both these cases the density matrix of
the neutrino would have no off-diagonal terms in the propagation eigenstate
basis.

Consider first the case when the flavour eigenstate $\nu_\alpha$ produced at
$t_0$ coincides with one of the instantaneous propagation eigenstates, which we
take to be $\nu_1$ for definiteness, i.e.\
$U_{\alpha i}(t_0)=\delta_{\alpha 1}$.
Then eqs.~(\ref{eq:rho10}) and ~(\ref{eq:S2a}) yield, for $t>t_1$,
\be
\rho_{12}(t)=
\int\frac{d^3p}{(2\pi)^3}
e^{-i\int_{t_1}^t[E_1(t',\vec{p})-E_2(t',\vec{p})]dt'}
|f_{\vec{p}_1}(\vec{p})|^2 \sin\alpha \cos\alpha\,.
\label{eq:rho11}
\ee
Next, we assume that the produced flavour eigenstates do not coincide with
one of the propagation eigenstates, but the production time $t_0$ of different
neutrinos is different, and averaging over it results in the vanishing of
the phase factors depending on $t_0$. Then we obtain from
(\ref{eq:rho10})
\be
\rho_{12}(t)=
\int\frac{d^3p}{(2\pi)^3}
e^{-i\int_{t_1}^t[E_1(t',\vec{p})-E_2(t',\vec{p})]dt'}
\big[|f_{\vec{p}_1}(\vec{p})|^2\overline{\cos^2\theta(t_0)}
-|f_{\vec{p}_2}(\vec{p})|^2
\overline{\sin^2\theta(t_0)}\big]\sin\alpha \cos\alpha\,,
\label{eq:rho12}
\ee
where we have assumed $\nu_\alpha=\nu_e$ and used $|U_{e 1}(t_0)|^2=\cos^2
\theta(t_0)$, $|U_{e 2}(t_0)|^2=\sin^2\theta(t_0)$. The bars over
$\sin^2(\theta_0)$ and $\cos^2(\theta_0)$ in (\ref{eq:rho12}) denote
averaging of these quantities over the production time $t_0$.

Let us now discuss eqs.~(\ref{eq:rho11}) and (\ref{eq:rho12}). We can see that
in both considered cases
($U_{\alpha i}(t_0)=\delta_{\alpha i}$ or averaging over $t_0$) the
off-diagonal elements of $\rho_{jk}(t)$ are non-zero. Moreover,
the integration in the exponents in eqs.~(\ref{eq:rho11})
and (\ref{eq:rho12}) is over the time
interval $[t_1, t]$ rather than over the full time interval $[t_0, t]$.
This means that in both cases the off-diagonal elements of $\rho_{jk}(t)$
do not depend on the evolution of the system before the time $t_1$. It can
also be readily seen from eqs.~(\ref{eq:rho11}) and (\ref{eq:rho12}) that
these elements
should exhibit
the usual damping with time due to WP separation, but the damping will
become important only for $t-t_1\gtrsim L_{\rm coh}$ rather than for
$t-t_0\gtrsim L_{\rm coh}$.

Thus, we have demonstrated that in the non-adiabatic regime
the off-diagonal elements of the density matrix in the propagation
eigenstate basis may be present and be damped with time due to WP separation
even when no such off-diagonal elements are initially present (and therefore
no WP separation is operative)
in the case of pure adiabatic evolution.

\subsubsection{\label{sec:model2}A model with extended adiabaticity violation}

The matrix $S_{ji}(t,t_0;\vec{p})$ can also be readily found in a model in
which non-adiabatic transitions between propagation eigenstates proceed at a
near constant pace. 
Let us elaborate on this. 
Let $\tau$ be the characteristic time over which the 
probability for the propagation eigenstates $\nu_1$ and $\nu_2$ to convert 
into each other takes its maximum value allowed by the degree of adiabaticity 
violation. We will be assuming that the matrix elements of the neutrino 
Hamiltonian in the propagation eigenstate basis $\tilde{\cal H}$~(\ref{eq:H3}) 
change very little during this time interval. In this case we 
can consider them as approximately time-independent. Up to an 
irrelevant overall phase factor, we then obtain for the evolution matrix 
$S_{ji}(t,t_0;\vec{p})$
\be
S(t,t_0;\vec{p})=\left(\begin{array}{cc}
\cos\Omega t +i\frac{\Delta-c_2\dot{\phi}}{2\Omega}\sin\Omega t
& ~~~-\frac{2\dot{\theta}+is_2\dot{\phi}} {2\Omega}\sin\Omega t
\vspace*{2.0mm}
\\
\frac{2\dot{\theta}-is_2\dot{\phi}}{2\Omega}\sin\Omega t &
~~~\cos\Omega t - i\frac{\Delta-c_2\dot{\phi}}{2\Omega}\sin\Omega t
\end{array}\right),
\label{eq:S3}
\ee
where
\be
2\Omega \equiv \sqrt{(\Delta-c_2\dot{\phi})^2+s_2^2\dot{\phi}^2+
4\dot{\theta}^2}\,.
\label{eq:Omega}
\ee
In the considered case, transitions between the propagation
eigenstates are oscillatory; the characteristic time $\tau$
mentioned above is then just half the oscillation period,
$\tau=\pi/(2\Omega)$.
The maximal amplitude of the
$\nu_1\leftrightarrow \nu_2$ transitions is $|2\dot{\theta}\pm is_2\dot{\phi}|/
(2\Omega)=\lambda/\sqrt{1+\lambda^2}$, with $\lambda$ being the
non-adiabaticity parameter defined in eq.~(\ref{eq:lambda}). Note that in the 
adiabatic limit $\dot{\theta}=\dot{\phi}=0$ eqs.~(\ref{eq:S3}) and 
(\ref{eq:Omega}) yield $S_{ji}(t,t_0;\vec{p})=\diag(e^{i\frac{\Delta}{2}t},\,
e^{-i\frac{\Delta}{2}t})$, in agreement with the results of 
Section~\ref{sec:adiab}.

Let us now assume that the initially produced neutrino is a $\nu_e$ and
consider the general expression for the
element $\rho_{12}(t)$ of the density matrix in the 2-flavour case. From
eqs.~(\ref{eq:rho10}) and (\ref{eq:U1}) we have
\begin{align}
\rho_{12}(t)=
\int\frac{d^3p}{(2\pi)^3}\big\{
&S_{11}S_{21}^*|f_{\vec{p}_1}(\vec{p})|^2 c^2(t_0)+
S_{11}S_{22}^*f_{\vec{p}_1}(\vec{p})f^*_{\vec{p}_2}(\vec{p})
s(t_0)c(t_0) \nonumber \\
+&S_{12}S_{21}^*f^*_{\vec{p}_1}(\vec{p})f_{\vec{p}_2}(\vec{p})
s(t_0)c(t_0)+S_{12}S_{22}^*|f_{\vec{p}_2}(\vec{p})|^2 s^2(t_0)\big\}\,.
\end{align}
Here we have taken into account that, since the neutrinos are produced at
$t=t_0$ as flavour eigenstates, the phase $\phi(t_0)$ vanishes.} Note that in
the adiabatic limit the evolution matrix $S_{ji}(t,t_0;\vec{p})$ is diagonal,
and only the second term in the curly brackets survives.

We shall now evaluate eq.~(\ref{eq:rho10a}) with the model (\ref{eq:S3}) for
the evolution matrix $S(t,t_0; \vec{p})$.  It will be convenient for us to
introduce the angle $\tilde{\theta}$ which characterizes the mixing of the
propagation eigenstates, and thus the adiabaticity violation.  The
non-adiabaticity parameter $\lambda$ introduced in eq.~(\ref{eq:lambda})
coincides with $\tan 2\tilde{\theta}$.%
\footnote{The mixing angle of the
propagation eigenstates $\tilde{\theta}$ must not be confused with the angle
$\theta$ describing flavour mixing in medium or its in-vacuum analogue
$\theta_0$.} 
Denoting $\tilde{s}_2\equiv\sin 2\tilde{\theta}$,
$\tilde{c}_2\equiv\cos 2\tilde{\theta}$, $\varphi\equiv \text{arg}[\tilde{\cal
H}_{12}(t)]$, from eq.~(\ref{eq:H3}) we find
\be
\tilde{s}_2=\frac{\sqrt{s_2^2\dot{\phi}^2+4\dot{\theta}^2}}{2\Omega}=
\frac{\lambda}{\sqrt{1+\lambda^2}}\,,\qquad
\tilde{c}_2=\frac{{\Delta-c_2\dot{\phi}}}{2\Omega}=
\frac{1}{\sqrt{1+\lambda^2}}\,,\qquad
\tan\varphi=-\frac{2\dot{\theta}}{s_2\dot{\phi}}\,.
\label{eq:propnot}
\ee
With this notation, we can rewrite eq.~(\ref{eq:S3}) as
\be
S_{ji}(t,t_0;\vec{p})=\left(\begin{array}{cc}
c_\Phi +i\tilde{c}_2 s_\Phi
& ~~~
-ie^{i\varphi}\tilde{s}_2 s_\Phi
\vspace*{2.0mm}
\\
-ie^{-i\varphi}\tilde{s}_2 s_\Phi &
~~~~~ c_\Phi - i\tilde{c}_2 s_\Phi
\end{array}\right),
\label{eq:S3a}
\ee
where
\be
c_\Phi\equiv\cos\Phi\,,\qquad
s_\Phi\equiv\sin\Phi\,,\qquad
\Phi\equiv\Omega (t-t_0)\,.
\label{eq:not3}
\ee
For the off-diagonal element $\rho_{12}(t)$ of the density matrix
in the propagation-eigenstate basis from eq.~(\ref{eq:rho10a}) we then find
\bea
\hspace*{-0.3cm}
\rho_{12}(t)\!=\!\int\frac{d^3p}{(2\pi)^3}|f_{\vec{p}_1}(\vec{p})|^2
\big\{c_\Phi^2 s(t_0) c(t_0)-s_\Phi^2 \big[\tilde{s}_2\tilde{c}_2
e^{i\varphi}[c^2(t_0)-s^2(t_0)]+(\tilde{c}_2^2-e^{2i\varphi}\tilde{s}_2^2)
s(t_0) c(t_0)\big]
\hspace*{-1.0cm}
\nonumber \\
+\,i s_\Phi c_\Phi\big[\tilde{s}_2 e^{i\varphi}[c^2(t_0)-s^2(t_0)]
+ 2\tilde{c}_2 s(t_0) c(t_0)\big]\big\},
\label{eq:rho13}
\eea
where for simplicity we neglected the small difference between 
the momentum distribution amplitudes of the WPs of the two propagation 
eigenstates $f_{\vec{p}_2}(\vec{p})$ and $f_{\vec{p}_1}(\vec{p})$.

Next, we note that our current assumption of near-constant elements of
$\tilde{\cal H}(t)$ in eq.~(\ref{eq:H3})
actually implies that $c\equiv \cos \theta$ and $s\equiv \sin \theta$
are practically constant, so that $\dot{\theta}\simeq 0$.
Eq.~(\ref{eq:propnot}) then gives $\varphi\simeq 0$, and from
eqs.~(\ref{eq:rho10a}) and~(\ref{eq:S3a}) we get
\bea
\rho_{12}(t)=\int\frac{d^3p}{(2\pi)^3}|f_{\vec{p}_1}(\vec{p})|^2
\big\{c_\Phi^2 s(t_0) c(t_0) - s_\Phi^2 \big[\tilde{s}_2\tilde{c}_2
[c^2(t_0)-s^2(t_0)]+(\tilde{c}_2^2-\tilde{s}_2^2) s(t_0) c(t_0)\big]
\nonumber \\
+ \,i s_\Phi c_\Phi\big[\tilde{s}_2 [c^2(t_0)-s^2(t_0)]
+ 2\tilde{c}_2 s(t_0) c(t_0)\big]\big\}.\quad
\label{eq:rho13a}
\eea
Notice that the assumption $\varphi\simeq 0$ adopted here actually implies
$|\dot{\theta}|\ll s_2|\dot{\phi}|$, which is opposite to the condition used
in Section~\ref{sec:model1}. Thus, the model in this subsection is in a sense
complementary to the one in Section~\ref{sec:model1}.

We are interested in the question of whether the off-diagonal elements of the
neutrino density matrix in the propagation eigenstate basis get suppressed at
late times. Consider therefore the regime of very large times, for which the
elements of $\rho_{jk}$ take time-independent asymptotic values.  It can be 
seen from eq.~(\ref{eq:rho13a}) that this corresponds to the averaging of the
oscillating terms in the integrand of the momentum integral.

Indeed, the width of the momentum interval that contributes significantly to
the integral in (\ref{eq:rho13a}) is given by the width $\sigma_p$ of the
momentum distribution functions $f_{\vec{p}_i}(\vec{p})$. If the phase
$\Phi=\Omega (t-t_0)$ of the oscillating factors in (\ref{eq:rho13a}) changes
significantly over this interval, i.e.\ if
\be
|\partial \Omega/\partial\vec{p}\,|\sigma_p (t-t_0) \gg 1\,,
\label{eq:coh2}
\ee
the $\Phi$-dependent terms in (\ref{eq:rho13a}) exhibit fast oscillations
and can be replaced by their average values.
Therefore the characteristic time after which the asymptotic regime sets in is
\be
t_{\rm char}\sim |\partial \Omega/\partial\vec{p}\,|^{-1}\sigma_x \,,
\label{eq:tchar}
\ee
where $\sigma_x\sim \sigma_p^{-1}$ is the spatial length of the neutrino wave
packet. In the adiabatic regime we have $2\Omega = \Delta$, and $t_{\rm char}$
essentially coincides with the adiabatic coherence length
$(L_{\rm coh})_{\rm adiab}\equiv |\partial \Delta/\partial\vec{p}\,|^{-1}
\sigma_x$; however, in the non-adiabatic case the characteristic time
$t_{\rm char}$ may differ significantly from the naive adiabatic coherence length
$(L_{\rm coh})_{\rm adiab}$. 

Consider now $\rho_{12}$ at asymptotically large times.
Averaging the $\Phi$-dependent oscillating factors, from
eq.~(\ref{eq:rho13a})) we find
\be
(\rho_{12})_{as}=\int\frac{d^3p}{(2\pi)^3}|f_{\vec{p}_1}(\vec{p})|^2
\Big\{\tilde{s}_2^2\,s(t_0) c(t_0)
- \frac{1}{2}\tilde{s}_2\tilde{c}_2
[c^2(t_0)-s^2(t_0)]
\Big\}.
\label{eq:rho14a}
\ee
Let us consider this expression in some limiting cases.
In the adiabatic limit $\lambda\to 0$ we have $\tilde{s}_2\to 0$ (no mixing
between the propagation eigenstates) and, as expected,  $\rho_{12}$ vanishes at
asymptotically large times due to the WP separation. In the opposite, extremely
non-adiabatic, limit $\lambda\to\infty$, we have $\tilde{s}_2=1$, which
corresponds to maximal mixing between the propagation eigenstates. In this
case eq.~(\ref{eq:rho14a}) yields $(\rho_{12})_{as}=s(t_0) c(t_0)$, i.e.\
the asymptotic value of $\rho_{12}(t)$ coincides with its initial value at
$t=t_0$. The overlap of the WPs of the different propagation eigenstates
does not reduce with time, i.e.\ no decoherence by WP separation occurs.
In the intermediate case of
a moderate adiabaticity violation, partial decoherence takes place.

These results can be readily understood.
In the adiabatic limit ($\lambda\to 0$), neutrino propagation eigenstates
evolve independently and propagate with well defined group velocities, which
are different for different eigenstates. Therefore, at asymptotically large
times their overlap essentially vanishes, leading to vanishing off-diagonal
elements of the neutrino density  matrix and complete decoherence.

In the opposite (extremely non-adiabatic) limit $\lambda\gg 1$, the 
propagation eigenstates, though mathematically well defined at any given $t$,
are strongly mixed and therefore are not a physically meaningful notion.  Their
group velocities are then not well defined either, and therefore no WP
separation can occur. The WP describing the initially produced neutrino flavour
eigenstate propagates as a single conglomerate, without separating into its
different propagation-eigenstate components. 
To put it another way, on a time scale $\tau=\pi/(2\Omega)$ which is short
compared to the coherence length expected in the adiabatic regime, the 
propagation eigenstates convert into each other with maximal amplitude. Each 
swap interchanges the WPs and so flips the sign of their group velocity 
difference. As a result, the (small) separation of the WPs over one period 
$\tau$ is completely compensated after the next such period, and no noticeable 
wave packet separation ever occurs.  Absence of WP separation in the extremely
non-adiabatic regime means that no decoherence takes place. The off-diagonal
elements of the neutrino density matrix at asymptotic times coincide with their
initial values at $t=t_0$, in agreement with the conclusions drawn above from
eq.~(\ref{eq:rho14a}).

In the case of moderate adiabaticity violation ($\lambda\sim 1$), the mixing
between the different propagation eigenstates is sizeable but not maximal.
The shuffling of the propagation eigenstates still occurs, but with an amplitude
smaller than one. Therefore, only a fraction of the faster eigenstate converts 
into the slower one and vice versa. As a result, at asymptotically large
times the separation of WPs still occurs, but the strengths of the separated
WPs are uneven: the probability of finding a neutrino in one of them is
larger than in the other.

It should be stressed that the above discussion applies only to the case of
continuous adiabaticity violation considered here; in the case when
adiabaticity is strongly violated only during a short fraction
of the overall neutrino evolution time, the situation is different,
as illustrated by the sample model discussed in
Section~\ref{sec:model1}.

\section{\label{sec:disc}Summary and discussion}

We have considered the effects of decoherence by WP separation on collective
neutrino oscillations in dense neutrino backgrounds. In particular, we have
studied in detail the question of whether decoherence effects can be described
by an extra term in the evolution equation of the density matrix of a neutrino
as a whole (as contrasted to the EoM of fixed-momentum components of the
density matrix) or of the neutrino density matrix in coordinate space.

Propagation decoherence occurs when the WP separation distance exceeds their 
length. An important quantity characterizing 
decoherence effects is therefore the length of WPs of neutrinos produced 
e.g.\ in supernovae, where collective neutrino oscillations are expected 
to occur. We considered this issue in Section~\ref{sec:size}. Our 
estimates show that, for neutrinos produced in the processes with 
participation of nucleons (such as $\beta$-processes or neutrino 
bremsstrahlung in nucleon collisions), the length $\sigma_x$ of the 
neutrino WPs is dominated by the overlap time of the nucleon WPs and is 
typically \mbox{$\lesssim 4\times 10^{-12}$ cm}. Interestingly, this 
estimate is rather close to the value $\sigma_x\sim 10^{-11}$ cm 
obtained in \cite{Kersten,KerstSmir} from completely different 
considerations, based on the assumption that for all production 
processes with participation of electrons $\sigma_x$ is dominated by the 
length of the electron WP.

In Section~\ref{sec:vac} we derived EoMs satisfied by both the
coarse grained (coordinate-averaged)
and un-averaged neutrino density matrices in vacuum. We
have found  that these equations essentially coincide, provided that in the
un-averaged case the total time derivative is understood as the Liouville
operator, with the average group velocity of the WPs of the neutrino mass
eigenstates playing the role of the neutrino transport velocity. Thus, modulo
this identification, the system is described by a Liouville-type equation
despite the fact that neutrinos are described by quantum WPs rather than being
considered as classical pointlike particles.

An important feature of the EoMs of the considered neutrino density matrices
in vacuum is that these equations contain a damping term, which 
leads to suppression with time of the off-diagonal elements of the neutrino 
density matrix in the mass eigenstate basis. This suppression is a consequence 
of the spatial separation of the WPs of the different neutrino mass eigenstates
composing the produced neutrino flavour state.
The presence of a term directly describing WP separation in the EoMs of the
coordinate-space or coordinate-averaged neutrino density matrices simplifies
studies of propagation decoherence, as in this case it is not necessary
to explicitly sum over neutrino momentum modes to check if their dephasing
occurs at late times.

Equipped with these results, we then studied the EoMs for the density matrices
describing neutrino flavour transitions in dense matter and neutrino
backgrounds, both in the adiabatic and in the non-adiabatic regimes.
We have shown that, just like for vacuum oscillations, in the adiabatic regime
the EoM of the neutrino density matrix contains a damping term which describes
decoherence by WP separation at late times. However, no such term in general
appears in the non-adiabatic regime, and whether or not propagation decoherence
occurs depends on the properties of the system. We 
have considered two specific models of adiabaticity violation -- one with 
strong short-term non-adiabaticity and another with extended non-adiabaticity 
of an arbitrary strength. For the first model, we found that WP separation 
does occur, but the degree of decoherence depends on the time elapsed after 
the short period of adiabaticity violation rather than on the overall time 
elapsed since neutrino production. For the second model, whether or not WP 
separation occurs depends on the extent of adiabaticity violation.

The obtained results allow us to give a quantitative interpretation of the well
known result \cite{RaffTamb,AM} that synchronized oscillations that can occur
in dense uniform and isotropic neutrino gases are not affected by decoherence
effects provided that the neutrino `self-interaction' parameter
$\mu=\sqrt{2}G_F n_\nu$ is sufficiently large (i.e.\ the neutrino density is
high enough). The interpretation depends on the value of $\mu P_0$ relative
to the mean `energy' of the neutrino WP, $\omega_0=[\Delta m^2/(2p)]_0$, and
its mean width in the $\omega$ variable, $\sigma_\omega$ (recall that $P_0$ is
the initial length of the global flavour spin vector).
As is shown in the Appendix, for
\be
\omega_0\ll \mu P_0\,
\label{eq:cond1}
\ee
we have $\lambda \sim s_{20}\omega_0/(\mu P_0)\ll 1$,
i.e. the neutrino system evolves adiabatically. In this case one could in
principle expect decoherence by WP separation, which could destroy the
synchronized oscillations. This, however, does not happen.
Indeed, eq.~(\ref{eq:angles2}) means that at the neutrino production
time $t_0$ one has $s_2(t_0)\simeq s_{20}\omega_0/(\mu P_0)\ll 
1$, that is the mixing in medium is suppressed, and the produced neutrino 
flavour eigenstate coincides with one of the propagation eigenstates. The
produced state therefore consists of just one WP, which cannot `separate
with itself'. As a result, no decoherence occurs.

If $\mu P_0$ satisfies
\be
\mu P_0 \sim \omega_0
\label{eq:cond2}
\ee
the non-adiabaticity parameter $\lambda$ is typically large (provided that
$\sigma_\omega \ll \omega_0$, so that from (\ref{eq:cond2}) it follows that
$\mu P_0 \sim \omega$ for all $\omega$ in the neutrino spectrum).
This means that the propagation eigenstates are strongly mixed, $\tilde{s}_2
\simeq 1$. It then follows from eq.~(\ref{eq:rho14a}) that the WP overlap at
late times is the same as it is at $t_0$, i.e.\ no decoherence by WP separation
takes place. The reason for this was discussed in Section~\ref{sec:nonadiab}.
Strong mixing between the propagation eigenstates means that they are not
physically meaningful quantities, and neither are their group velocities;
therefore there is no WP separation. Another way of saying this is that
the propagation eigenstates go into each other (shuffle) on a very short time
scale $\tau$, which precludes any observable WP separation.
Thus, coherence is maintained in that case. If, however, 
$\sigma_\omega$ is comparable to $\omega_0$, eq.~(\ref{eq:cond2}) does not 
guarantee that $\lambda$ is large for all $\omega$-modes, and partial or full 
decoherence by WP separation takes place at late times, leading to partial
or complete de-synchronization of neutrino oscillations.

It has been shown in \cite{RaffTamb,AM} that synchronized 
neutrino oscillations can occur (and so no decoherence by WP separation 
takes place) also in the limit
\be
\sigma_\omega\ll \mu P_0 \ll \omega_0\,. 
\label{eq:cond3} 
\ee
In this case $\lambda\sim s_{20}\mu P_0/\omega_0\ll 1$, i.e.\ adiabaticity is 
good, and mixing at neutrino production essentially coincides 
with vacuum mixing. Therefore, the initially produced neutrino flavour state 
is again a nontrivial mixture of different propagation eigenstates, 
and naively one could expect decoherence by WP separation and loss of 
synchronization at late times. It is not difficult to understand why 
this actually does not happen. Eq.~(\ref{eq:cond3}) means $\sigma_\omega \lll
\omega_0$, so that the neutrino spectrum is practically monochromatic. 
Therefore, 
all neutrinos oscillate with essentially the same frequency, i.e.\ experience 
synchronized oscillations.%
\footnote{One could still expect decoherence to take place in 
this case, but only at extremely late times $\sim \sigma_{\omega}^{-1}$.}
The absence of WP separation can also be readily understood at a formal level 
by considering the off-diagonal elements of the neutrino density matrix. 
In this case the spectrum shape function $f_{\vec{p}_i}(\vec{p})$, when transformed 
to the $\omega$-variable, is essentially $\delta(\omega-\omega_0)$, and 
no time averaging should be performed. Therefore, instead of 
eq.~(\ref{eq:rho14a}) one should use eq.~(\ref{eq:rho13a}) or~(\ref{eq:rho13}). 
Good adiabaticity means $\tilde{s}_2\to 0$, $\tilde{c}_2\to 1$, and 
eq.~(\ref{eq:rho13a}) (or~(\ref{eq:rho13})) yields
\be
\rho_{12}(t)=e^{2i\Phi} s(t_0) c(t_0)\,,
\label{eq:}
\ee
which up to the phase factor coincides with $\rho_{12}(t_0)=s(t_0) c(t_0)$. 

As shown in the Appendix, with $\mu P_0$ decreasing and approaching 
$\mu_0 P_0\sim \sigma_\omega$, 
the non-adiabaticity parameter $\lambda$ decreases, which leads to partial 
decoherence by WP separation at late times. For $\mu\le\mu_0$ adiabaticity 
becomes perfect ($\lambda=0$). In this regime the in-medium mixing at neutrino 
production 
is of the order of the vacuum mixing and therefore in general 
is not suppressed. Thus, unlike in the case (\ref{eq:cond1}), the 
initially produced neutrino flavour eigenstate is a nontrivial superposition 
of different propagation eigenstates. At sufficiently large times their WPs 
fully separate, destroying synchronized neutrino oscillations.

We have addressed only one aspect of late-time decoherence of oscillations in 
dense neutrino gases -- decoherence by WP separation. To this end, we 
considered a simplified model -- a uniform and isotropic gas  
of neutrinos described by identical WPs. In realistic situations, one 
should also take into account that the neutrino ensemble contains WPs with  
different centroid momenta and perform a statistical averaging over this 
ensemble, which will also contribute to decoherence effects.
Our evolution equations for the density matrix will take this into
account automatically, provided that by the momentum distribution functions
$|f(\vec{p})|^2$ we understand the overall momentum spectrum (that 
is, the sum of the spectra of all the individual WPs). Then equations like our 
eq.~(\ref{eq:EoM1m}) (in the adiabatic regime) or (\ref{eq:EoM3m}), 
(\ref{eq:EoM4m}) (in the general case) will describe the space-time evolution 
of the system and the loss of coherence in it. In particular, some distance or 
time (coherence time) after which the coherence is lost can be found. However, 
they in general cannot be interpreted as the distance or time 
after which the neutrino WPs separate. Thus, interpretation in terms of WP 
separation will be lost in this case, or 
will only be a part of the story.

In conclusion, the results of this paper provide novel analytical tools
for describing effects of decoherence by wave packet separation 
in neutrino oscillations.
In particular, in Sections~\ref{sec:vac} and \ref{sec:medium} we have
derived effective equations of motion which describe the evolution of
a neutrino wave packet as a whole, as opposed to the evolution of its
individual momentum modes.  These equations could be used, for instance,
in simulations of supernova neutrinos to incorporate decoherence effects
without the need to simulate a large number of momentum modes. 

\vspace*{1.5mm}
While this paper was in the stage of the manuscript preparation, the
paper \cite{HansSmi} appeared, in which, within the wave packet formalism, the
Liouville equation for the neutrino density matrix describing neutrino flavour
evolution in matter was derived. An extra term in the equation of motion 
responsible for decoherence due to wave packet separation was obtained and 
analyzed. However, the nonadiabatic neutrino evolution and the oscillations 
in neutrino backgrounds were not considered there. Where the results of the 
present paper and of \cite{HansSmi} overlap, they are in agreement with each 
other.

\vspace*{1.5mm}
{\em Acknowledgments.} The authors are grateful to Alexei Smirnov and Irene
Tamborra for useful discussions and to Antonino Di Piazza for pointing out a
typo in eq.~(\ref{eq:Lcoh02}).

\appendix
\renewcommand{\theequation}{\thesection\arabic{equation}}
\appsection
\renewcommand{\thesection}{\Alph{section}}
\section*{Appendix \Alph{section}:
The flavour spin formalism and the adiabaticity\\
\hspace*{3.3cm}
condition
}

Consider a uniform and isotropic neutrino gas in the 2-flavour framework.  The
evolution of such a system is conveniently described within the flavour spin
formalism \cite{SR}.  For an isotropic system, one can use the absolute value
of the neutrino momentum $p\equiv |\vec{p}|$ rather than the momentum itself in
order to label the kinematic characteristics of neutrinos.  It proves to be
more convenient, however, to use instead the vacuum oscillation frequency
$\omega\equiv \Delta m^2/(2p)$. The $2\times 2$ density matrix in flavour
space can be expanded in terms of the unit matrix and Pauli matrices as
\be
\rho_\omega=
\frac{n_{\nu\omega}}{2}
\big(P_\omega^0+\vec{\sigma}\vec{P}_\omega\big)\,.
\label{eq:P1}
\ee
Here $n_{\nu\omega}$ is the number density of the $\omega$-mode neutrinos, and
$\vec{P}_\omega$ is the corresponding flavour spin vector.
The 2-flavour effective neutrino Hamiltonian can be similarly expanded as
\be
{\cal H}(t)=\frac{1}{2}\big(H_{0\omega}+\vec{\sigma}\vec{H}_\omega\big)\,,
\ee
where the ``Hamiltonian vector'' $\vec{H}_\omega$ is given by
\be
\vec{H}_\omega=\omega\vec{B}+\mu \vec{P}\,.
\label{eq:vecH1}
\ee
Here
\be
\vec{B}=(s_{20},\,0,\,-c_{20})\,,\qquad \mu=\sqrt{2}G_F n_\nu\,,
\label{eq:omegaBmu}
\ee
where, as discussed in Section~\ref{sec:size}, $s_{20}\equiv\sin 2\theta_0$,
$c_{20}\equiv\cos 2\theta_0$, with $\theta_0$ being the leptonic mixing angle
in vacuum, and $n_\nu$ is the total density of neutrinos. The quantity
$\vec{P}$ is the global flavour spin vector which is the sum (integral) of the
individual flavour spin vectors $\vec{P}_\omega$ corresponding to fixed
$\omega$:
\be
\vec{P}=\int_{-\infty}^\infty d\omega \vec{P}_\omega\,.
\label{eq:sum}
\ee
Here positive and negative $\omega$ correspond to neutrinos and antineutrinos,
respectively. The terms in $\rho_\omega$ and ${\cal H}$ that are proportional
to the unit matrix do not affect neutrino evolution and will hereafter be
omitted, i.e.\ we will consider only the traceless parts of
$\rho_\omega$ and ${\cal H}$.

The evolution equation for the density matrix (\ref{eq:SR}) yields the
following EoM for the individual flavour spin vectors:
\be
\dot{\vec{P}}_\omega=\vec{H}_\omega\times\vec{P}_\omega\,,
\label{eq:EoMPomega}
\ee
i.e.\ the vectors $\vec{P}_\omega$ precess around their corresponding
$\vec{H}_\omega$ with frequencies $H_\omega\equiv |\vec{H}_\omega|$.
Integrating eq.~(\ref{eq:EoMPomega}) over $\omega$, we obtain the EoM for the
global flavour spin vector:
\be
\dot{\vec{P}}=\vec{B}\times\vec{S}\,,\qquad{\rm where}
\qquad \vec{S}\equiv \int d\omega \, \omega \vec{P}_\omega\,.
\label{eq:EoMP}
\ee
From EoMs (\ref{eq:EoMPomega}) and (\ref{eq:EoMP}) it follows that the
quantities $|\vec{P}_\omega|$ and $\vec{B}\vec{P}$ are conserved by neutrino
evolution.

The flavour content of a given $\omega$-mode is defined by the projection of
$\vec{P}_\omega$ onto the 3rd axis in flavour space.
We will be assuming that all neutrinos are produced at $t=0$ in the electron
flavour, which means that the initial conditions for $\vec{P}_\omega$ and
$\vec{P}$ are
\be
\vec{P}_\omega(0)=P_0 g_\omega \vec{n}_3\,, \qquad \vec{P}(0)=P_0\vec{n}_3\,,
\label{eq:init}
\ee
where $g_\omega$ is the neutrino spectrum in the $\omega$ variable normalized
according to $\int \, d\omega g_\omega=1$, and $\vec{n}_3$ is the unit vector in
the third direction in flavour space.  Then $|\vec{P}_\omega|=P_0 g_\omega$,
$\vec{B} \vec{P}=-c_{20}P_0$, and the length of the vector $\vec{H}_\omega$
is given by
\be
H_\omega=
\big\{(\omega-\mu P_0 c_{20})^2+\mu^2(P^2-c_{20}^2 P_0^2)\big\}^{1/2}\;.
\label{eq:length}
\ee
The traceless Hermitian $2\times 2$ Hamiltonian matrix ${\cal H}(t)$ can be
written in terms of the components of the real
vector $\vec{H}_\omega=(H_{\omega 1},\,H_{\omega 2},\,H_{\omega 3})$ as
\be
{\cal H}(t)=\frac{1}{2}\vec{\sigma}\vec{H}_\omega
=\frac{1}{2}
\left(\begin{array}{cc}
H_{\omega 3} & H_{\omega1}\!-\!iH_{\omega 2}  \\
H_{\omega 1}\!+\!iH_{\omega 2}& -H_{\omega 3}  \end{array} \right).
\label{eq:H2mm}
\ee
Comparing eqs.~(\ref{eq:H2mm}) and (\ref{eq:H1mm}), one can find the quantities
$c_2\equiv\cos 2\theta(t)$, $s_2\equiv\sin 2\theta(t)$
and $\phi(t)$, through which the Hamiltonian ${\cal H}$ in eq.~(\ref{eq:H1mm})
is expressed, in terms of the components of the
Hamiltonian vector $\vec{H}_\omega$
and $\Delta\equiv E_2(t)-E_1(t)$:
\be
c_2=-\frac{H_{\omega 3}}{\Delta}
=\frac{c_{20}\omega-\mu P_3}{\Delta}\,,\qquad\quad
s_2=\frac{H_{\omega\perp}}{\Delta}
\,,\qquad\quad
\tan\phi=-\frac{H_{\omega 2}}{H_{\omega 1}}\,.
\quad
\label{eq:angles2}
\ee
Here we used the notation $H_{\omega\perp}\equiv\sqrt{H_{\omega 1}^2+
H_{\omega 2}^2}$. Note that
\be
|\vec{H}_\omega| \equiv H_\omega=|\Delta|\,.
\label{eq:Delta}
\ee
The angles $\theta(t)$ and $\phi(t)$ have simple interpretation in
the flavour spin formalism: $2\theta$ is the angle between the Hamiltonian
vector $\vec{H}_\omega$ and the 3rd direction in
flavour space, i.e.\ it
is the polar angle of $\vec{H}_\omega$, whereas $\phi(t)$ is
the azimuthal angle characterizing the direction of the projection of
$\vec{H}_\omega$ onto the 1--2 plane.

The non-adiabaticity parameter $\lambda$ defined in eq.~(\ref{eq:lambda})
depends on the derivatives $\dot{\theta}$ and $\dot{\phi}$.
Let us express them in terms of the components of $\vec{H}_\omega$ and their
derivatives. From the definitions in eq.~(\ref{eq:angles2}) we find
\bea
&&\dot{\phi}=\frac{\dot{H}_{\omega 1} H_{\omega 2}-\dot{H}_{\omega 2}
H_{\omega 1}}{H_{\omega\perp}^2}=
\frac{(\dot{\vec{H}}_{\omega}\times \vec{H}_{\omega})_3}{H_{\omega\perp}^2}\,,
\label{eq:deriv1}
\\
&&2\dot{\theta}=\frac{\dot{H}_{\omega 3} H_{\omega\perp}-
\dot{H}_{\omega\perp} H_{\omega 3}}{H_\omega^2}=
\frac{\dot{H}_{\omega 3} H_\omega^2-H_{\omega 3}(\dot{\vec{H}}_\omega
\vec{H}_\omega)}{H_\omega^2 H_{\omega\perp}}=
\frac{[\vec{H}_\omega\times(\dot{\vec{H}}_\omega\times\vec{H}_\omega)]_3}
{H_\omega^2 H_{\omega\perp}}\,.~~~~~
\label{eq:deriv2}
\eea
It proves convenient to introduce the angular velocity of the vector
$\vec{H}_\omega$, i.e.\ the derivative of the unit vector in its direction:
\be
\vec{\Omega}_{H_\omega}\equiv\frac{d}{dt}\frac{\vec{H}_\omega}
{H_\omega}=\frac{\vec{H}_\omega\times(\dot{\vec{H}}_\omega\times
\vec{H}_\omega)}{{H}_\omega^3}\,.
\label{eq:OmegaH}
\ee
Eq.~(\ref{eq:deriv2}) then takes the form
$2\dot{\theta}=(H_\omega/H_{\omega\perp})(\Omega_{H_\omega})_3$. It is not
difficult to show that
the expression under the square root in the numerator of the formula
(\ref{eq:lambda}) for $\lambda$
is just the squared length of $\vec{\Omega}_{H_\omega}$:
\be
s_2^2\dot{\phi}^2+4\dot{\theta}^2=
\frac{(\dot{\vec{H}}_\omega
\times \vec{H}_\omega)^2}{H_\omega^4}=\vec{\Omega}_{H_\omega}^2.
\label{eq:exp1}
\ee
The non-adiabaticity parameter $\lambda$ can therefore be written as
\be
\lambda =\frac{\Omega_{H_\omega}}
{|\Delta -c_2\dot{\phi}|}\,.
\label{eq:lambda2}
\ee
Note that it is similar (but not identical) to the non-adiabaticity parameter
\be
\frac{\Omega_{H_\omega}}{H_\omega}=\frac{\Omega_{H_\omega}}{|\Delta|}
\label{eq:nonadDuan}
\ee
introduced in \cite{duan1}.  The difference between the two parameters is due
to the fact that they have somewhat different meaning. The one in
eq.~(\ref{eq:nonadDuan}) is the ratio of the speed $\Omega_{H_\omega}$ with
which the direction of $\vec{H}_\omega$ changes to the angular velocity
$H_\omega$ of precession of $\vec{P}_\omega$ around $\vec{H}_\omega$: when this
ratio is small, the individual flavour spin vectors, while precessing around
their $\vec{H}_\omega$, track the motion of the latter. On the other hand, our
parameter $\lambda$ is defined as the tangent of twice the mixing angle between
the propagation eigenstates; it describes to what extent the independence of
the evolution of these eigenstates is violated. Both parameters characterize
the degree of adiabaticity violation in the neutrino system.

The evolution of the system under consideration exhibits a threshold
behaviour in the neutrino self-interaction parameter $\mu$:
there exists a critical value $\mu_0$ which depends on the neutrino spectrum
$g_\omega$ and is of the order of the width of this spectrum $\sigma_\omega$
\cite{RaffTamb,AM}.
For $\mu\le \mu_0$ complete decoherence is achieved at asymptotic (late) times,
whereas for $\mu>\mu_0$ the system maintains partial or complete coherence
at all times.
For $\mu>\mu_0$, the
late time evolution
of the global flavour spin vector $\vec{P}$ is to a good approximation a simple
precession around the vector $\vec{B}$ with a
frequency $\omega_s$:
\be
\dot{\vec{P}}\approx\omega_s \vec{B}\times\vec{P}\,.
\label{eq:EoM1as1}
\ee
In this regime the length of the vector $\vec{P}$ is (approximately) conserved;
however, it may change significantly at short and intermediate times, before
the asymptotic regime sets in.  For $\mu\gg \mu_0$ the EoM in
(\ref{eq:EoM1as1}) is practically exact, and the asymptotic regime described by
this equation sets in immediately at $t=0$; the length of the vector $P$ is
therefore preserved by neutrino evolution, i.e.\ $P=P_0$. The precession
frequency $\omega_s$ in this case coincides with the spectral average
$\omega_0\equiv \int g_\omega \omega \, d\omega$. The flavour spin vectors
$\vec{P}_\omega$ of all the individual modes precess around $\vec{B}$ with the
same frequency $\omega_0$ and in phase with each other, i.e.\ synchronized
oscillations take place.

For $\mu\gtrsim \mu_0$, i.e.\ $\mu$ exceeding $\mu_0$ but not by much, the
late-time EoM (\ref{eq:EoM1as1}) is only approximate; the length of the
component of $\vec{P}$ orthogonal to $\vec{B}$ is actually not exactly
conserved, but exhibits small oscillations. Still, averaging over these small
oscillations (which leads to eq.~(\ref{eq:EoM1as1})) is a good first
approximation. In this case the length of the vector $\vec{P}$ shrinks during
the time period before the asymptotic regime sets in, i.e.\ the asymptotic
value of $P$ satisfies $P_\text{min}< P< P_0$. The existence of a non-zero
minimum value $P_\text{min}=c_{20}P_0$ follows from the conservation of
$\vec{B} \vec{P}$.  The shrinkage of $P$ is a consequence of partial dephasing
of the individual $\omega$-modes. In configuration-space description, it is
related to decoherence by WP separation, which, however, is not complete in
this case (see Section~\ref{sec:model2}).  The precession frequency
$\omega_s$ is in general different from $\omega_0$.

For $\mu\le \mu_0$, at asymptotically large times the flavour spin vector
$\vec{P}$ aligns with $\vec{B}$ and its evolution stops. Its length $P$ takes
its minimum possible value $P=c_{20}P_0$, which means
complete dephasing of the individual $\omega$-modes, or equivalently complete
decoherence by WP separation in configuration space.

Let us now consider late-time evolution of $\vec{P}$ for $\mu\ge\mu_0$
assuming the EoM in eq.~(\ref{eq:EoM1as1}) to be exact.
It is not difficult to solve this equation. To fix the initial conditions,
we have to specify the value of $P=|\vec{P}|$, as well as the direction of the
vector $\vec{P}$ at a given time $t_0$.
Let us introduce the parameter $R$ describing the length of this vector
in the asymptotic regime:
\be
R=\sqrt{\frac{P^2}{P_0^2}-c_{20}^2}\,,\qquad
P^2=P_0^2\big(c_{20}^2+R^2\big)\,.
\label{eq:R}
\ee
The length of the
component of $\vec{P}$ that is orthogonal to $\vec{B}$ is then given by%
\footnote{The notation $\vec{P}_{\perp\vec{B}}$ was introduced here to stress
the difference between the vector components orthogonal to $\vec{B}$ from
those orthogonal to the 3rd axis in flavour space, cf.\ eq.~(\ref{eq:angles2}).
}
\be
P_{\perp\vec{B}}\equiv|\vec{P}_{\perp\vec{B}}|=P_0 R\,.
\label{eq:Pperp}
\ee
Note that the quantity $R/s_{20}$ can be considered as the order parameter
characterizing decoherence in the system. Indeed,
for $\mu\gg \mu_0$ we have $P\simeq P_0$ and $R\simeq s_{20}$,
whereas at the threshold $\mu=\mu_0$ we obtain $P=P_\text{min}=c_{20}P_0$,
$R=0$.

We shall adopt the initial condition such that at a time $t_0$ the vector
$\vec{P}$ lies in the 1--3 plane in flavour space. Direct 
integration of eq.~(\ref{eq:EoM1as1}) then yields
\bea
&&P_1=P_0 c_{20}(-s_{20} + R c_\xi)\,,
\vspace*{1.2mm}
\nonumber \\
&&P_2=-P_0 R s_\xi\,,
\vspace*{1.4mm}
\nonumber \\
&&P_3=P_0(c_{20}^2+s_{20} R c_\xi)\,,
\label{eq:sols1}
\eea
where
\be
c_\xi\equiv \cos\xi\,,\qquad
s_\xi\equiv \sin\xi\,,\qquad
\xi\equiv \omega_s (t-t_0)\,.
\label{eq:xi}
\ee

The components of the vector $\vec{H}_\omega$ can then be found by substituting
(\ref{eq:sols1}) into (\ref{eq:vecH1}):
\bea
&& H_{\omega 1}=s_{20}(\omega-c_{20}\mu P_0)+c_{20}\mu P_0 R c_\xi\,,
\vspace*{1.2mm}
\nonumber \\
&& H_{\omega 2}=-\mu P_0 R s_\xi
\vspace*{1.2mm}\,,
\nonumber \\
&& H_{\omega 3}=-c_{20}(\omega-c_{20}\mu P_0) + s_{20} \mu P_0 R c_\xi \,.
\label{eq:sols2}
\eea
It should be remembered that in the above expressions for the components of
$\vec{H}_\omega$ the frequency $\omega$ in general
does not coincide with $\omega_s$, though for narrow neutrino spectra
($\sigma_\omega \ll \omega_s$) they must be close to each other.

For the lengths of $\vec{H}_\omega$ and of its component
$\vec{H}_{\omega\perp}$ orthogonal to the 3rd axis in flavour space
we then find
\be
H_\omega=
\left\{(\omega-c_{20}\mu P_0)^2+\mu^2 P_0^2 R^2\right\}^{1/2},
\qquad\qquad\qquad\qquad~\,
\label{eq:H}
\ee
\be
H_{\omega\perp}=
\left\{[s_{20}(\omega-c_{20}\mu P_0)+c_{20}\mu P_0 R c_\xi]^2
+\mu^2 P_0^2 R^2 s_\xi^2\right\}^{1/2}.
\label{eq:Hperp}
\ee
Note that eq.~(\ref{eq:Pperp}) means that~(\ref{eq:H})
is consistent with eq.~(\ref{eq:length}).

The mixing angle in neutrino background $\theta(t)$ and the angle $\phi(t)$
characterizing the projection of $\vec{H}_\omega$ onto the 1--2 plane in 
flavour space are defined through the components of the vector $\vec{H}_\omega$
by the relations in eq.~(\ref{eq:angles2}). Substituting there the expressions
for these components from eq.~(\ref{eq:sols2}) we obtain, in the case when the
asymptotic evolution of $\vec{P}$ is described by eq.~(\ref{eq:EoM1as1}),
\be
2\dot{\theta}=\frac{\dot{H}_{\omega 3}}{H_{\omega\perp}}=
-\frac{\omega_s s_{20} \mu P_0 R s_\xi}
{\left\{[s_{20}(\omega-c_{20}\mu P_0)+c_{20}\mu P_0 R c_\xi]^2
+\mu^2 P_0^2 R^2 s_\xi^2 \right\}^{1/2}}\,.
\vspace*{1.5mm}
\label{eq:thetadot}
\ee
\be
\dot{\phi}=\frac{
\omega_s\mu P_0 R \left[s_{20}(\omega-c_{20}\mu P_0) c_\xi+c_{20}\mu P_0 R 
\right]}
{[s_{20}(\omega-c_{20}\mu P_0)+c_{20}\mu P_0 R c_\xi]^2
+\mu^2 P_0^2 R^2 s_\xi^2 
}\,.
\vspace*{0.8mm}
\label{eq:phidot}
\ee
In calculating $2\dot{\theta}$ we have taken into account that
eqs.~(\ref{eq:vecH1}) and (\ref{eq:EoM1as1}) imply that $\vec{H}_\omega$
satisfies $\dot{\vec{H}}_\omega=\omega_s\vec{B}\times\vec{H}_\omega$, i.e.\
in the regime described by EoM (\ref{eq:EoM1as1}), $\vec{H}_\omega^2=\text{const.}$
and $\dot{\vec{H}}_\omega\vec{H}_\omega=0$.
Note that, while $\dot{\theta}$ oscillates around zero or small
average, taking both positive and negative values, $\dot{\phi}$ oscillates
around a non-zero value, remaining for most of the parameter space
sign-definite.

Let us calculate the numerator of expression (\ref{eq:lambda}) for
$\lambda$ in the regime described by EoM (\ref{eq:EoM1as1}). Making use of
eqs.~(\ref{eq:phidot}) and (\ref{eq:thetadot}), we find
\be
\sqrt{s_2^2\dot{\phi}^2+4\dot{\theta}^2}=\Omega_{H_\omega}=
\frac{\omega_s \mu P_0 R}{H_\omega}\,,
\label{eq:numer}
\ee
which is time-independent. For the non-adiabaticity parameter $\lambda$ we
find
\be
\lambda=\frac{\omega_s \mu P_0 R}{H_\omega|\Delta-c_2\dot{\phi}|}
=\frac{\omega_s \mu P_0 R}{H_\omega|\pm H_\omega-c_2\dot{\phi}|}\,.
\label{eq:lambda3}
\ee
Consider first the limit $\mu P_0\gg \mu_0 P_0\sim \sigma_\omega$. In this case 
$P\to P_0$, $R\to s_{20}$, $\omega_s\to \omega_0$. From eqs.~(\ref{eq:phidot}), 
(\ref{eq:angles2}) and (\ref{eq:length}) it then follows that one can neglect 
$c_2 \dot{\phi}$ compared to $H_\omega$ in eq.~(\ref{eq:lambda3}) 
provided that $\mu P_0\gg \omega$ or $\mu P_0\ll \omega$. This gives 
$\lambda\sim s_{20}\omega_0\mu P_0/H_\omega^2$. For $\mu P_0\gg \omega_0$
we have $H_\omega\approx \mu P_0$ and $\lambda\sim s_{20}\omega_0/(\mu P_0)\ll 
1$, whereas for $\mu P_0\ll \omega_0$ we obtain $H_\omega\approx \omega_0$ and 
$\lambda\sim s_{20}\mu P_0/\omega_0 \ll 1$. Thus, 
in the limits when $\mu P_0$ is vastly different from $\omega_0$, neutrino 
evolution is adiabatic. 
If $\mu P_0 \sim \omega_0$, one cannot in general neglect 
$c_2\dot{\phi}$ compared to $H_\omega$ in the denominator in 
eq.~(\ref{eq:lambda3}). In this case 
$\lambda$ is either $\sim 1$ or $\gg 1$, i.e.\ adiabaticity is either 
moderately or strongly violated. 

For $\mu$ approaching $\mu_0$ from above, we have 
$R\to 0$, so that $\lambda\to 0$, and neutrino evolution at late times is 
fully adiabatic. The same is also true for $\mu < \mu_0$.


\begin{thebibliography}{99}

\bibitem{synch1}
  S.~Samuel,
  ``Neutrino oscillations in dense neutrino gases,''
  Phys.\ Rev.\ D {\bf 48}, 1462 (1993).

\bibitem{synch2}
  V.~A.~Kostelecky, J.~T.~Pantaleone and S.~Samuel,
  ``Neutrino oscillation in the early universe,''
  Phys.\ Lett.\ B {\bf 315}, 46 (1993).

\bibitem{Pantaleone:1992eq}
  J.~T.~Pantaleone,
  ``Neutrino oscillations at high densities,''
  Phys.\ Lett.\ B {\bf 287}, 128 (1992).

\bibitem{synch3}
  J.~T.~Pantaleone,
  ``Stability of incoherence in an isotropic gas of oscillating neutrinos,''
  Phys.\ Rev.\ D {\bf 58}, 073002 (1998).

\bibitem{synch4}
  S.~Pastor, G.~G.~Raffelt and D.~V.~Semikoz,
  ``Physics of synchronized neutrino oscillations caused by selfinteractions,''
  Phys.\ Rev.\ D {\bf 65}, 053011 (2002)
  [hep-ph/0109035].


\bibitem{synch5}
  A.~D.~Dolgov, S.~H.~Hansen, S.~Pastor, S.~T.~Petcov, G.~G.~Raffelt and D.~V.~Semikoz,
  ``Cosmological bounds on neutrino degeneracy improved by flavor
  oscillations,''
  Nucl.\ Phys.\ B {\bf 632} (2002) 363
  [hep-ph/0201287].

\bibitem{synch6}
  G.~M.~Fuller and Y.~-Z.~Qian,
  ``Simultaneous flavor transformation of neutrinos and antineutrinos with
  dominant potentials from neutrino-neutrino forward scattering,''
  Phys.\ Rev.\ D {\bf 73}, 023004 (2006)
  [astro-ph/0505240].

\bibitem{RaffTamb}
  G.~G.~Raffelt and I.~Tamborra,
  ``Synchronization versus decoherence of neutrino oscillations at
   intermediate densities,''
  Phys.\ Rev.\ D {\bf 82} (2010) 125004
  [arXiv:1006.0002 [hep-ph]].

\bibitem{bipolar}
  V.~A.~Kostelecky and S.~Samuel,
  ``Selfmaintained coherent oscillations in dense neutrino gases,''
  Phys.\ Rev.\ D {\bf 52}, 621 (1995)
  [hep-ph/9506262].


\bibitem{duan1}
  H.~Duan, G.~M.~Fuller and Y.~-Z.~Qian,
  ``Collective neutrino flavor transformation in supernovae,''
  Phys.\ Rev.\ D {\bf 74}, 123004 (2006)
  [astro-ph/0511275].

\bibitem{Duan:2006an}
  H.~Duan, G.~M.~Fuller, J.~Carlson and Y.~Z.~Qian,
  ``Simulation of Coherent Non-Linear Neutrino Flavor Transformation in the Supernova Environment. 1. Correlated Neutrino Trajectories,''
  Phys.\ Rev.\ D {\bf 74}, 105014 (2006)
  [astro-ph/0606616].

\bibitem{hann1}
  S.~Hannestad, G.~G.~Raffelt, G.~Sigl and Y.~Y.~Y.~Wong,
  ``Self-induced conversion in dense neutrino gases: Pendulum in flavour space,''
  Phys.\ Rev.\ D {\bf 74}, 105010 (2006)
  [Erratum-ibid.\ D {\bf 76}, 029901 (2007)]
  [astro-ph/0608695].

\bibitem{Fogli:2007bk}
  G.~L.~Fogli, E.~Lisi, A.~Marrone and A.~Mirizzi,
  ``Collective neutrino flavor transitions in supernovae and the role of trajectory averaging,''
  JCAP {\bf 0712}, 010 (2007)
  [arXiv:0707.1998 [hep-ph]].

\bibitem{duan2}
  H.~Duan, G.~M.~Fuller, J.~Carlson and Y.~-Z.~Qian,
  ``Analysis of Collective Neutrino Flavor Transformation in Supernovae,''
  Phys.\ Rev.\ D {\bf 75}, 125005 (2007)
  [astro-ph/0703776].

\bibitem{splits1}
  H.~Duan, G.~M.~Fuller, J.~Carlson and Y.~-Z.~Qian,
  ``Simulation of Coherent Non-Linear Neutrino Flavor Transformation in
  the Supernova Environment. 1. Correlated Neutrino Trajectories,''
  Phys.\ Rev.\ D {\bf 74}, 105014 (2006)
  [astro-ph/0606616].

\bibitem{splits2}
  G.~G.~Raffelt and A.~Y.~Smirnov,
  ``Self-induced spectral splits in supernova neutrino fluxes,''
  Phys.\ Rev.\ D {\bf 76}, 081301 (2007)
  [Erratum-ibid.\ D {\bf 77}, 029903 (2008)]
  [arXiv:0705.1830 [hep-ph]].

\bibitem{splits3}
  H.~Duan, G.~M.~Fuller, J.~Carlson and Y.~-Q.~Zhong,
  ``Neutrino Mass Hierarchy and Stepwise Spectral Swapping of Supernova
  Neutrino Flavors,''
  Phys.\ Rev.\ Lett.\  {\bf 99}, 241802 (2007)
  [arXiv:0707.0290 [astro-ph]].

\bibitem{splits4}
  G.~G.~Raffelt and A.~Y.~Smirnov,
  ``Adiabaticity and spectral splits in collective neutrino transformations,''
  Phys.\ Rev.\ D {\bf 76}, 125008 (2007)
  [arXiv:0709.4641 [hep-ph]].

\bibitem{multisplits}
  B.~Dasgupta, A.~Dighe, G.~G.~Raffelt and A.~Y.~Smirnov,
  ``Multiple Spectral Splits of Supernova Neutrinos,''
  Phys.\ Rev.\ Lett.\  {\bf 103}, 051105 (2009)
  [arXiv:0904.3542 [hep-ph]].

\bibitem{review1}
  H.~Duan, G.~M.~Fuller and Y.~-Z.~Qian,
  ``Collective Neutrino Oscillations,''
  Ann.\ Rev.\ Nucl.\ Part.\ Sci.\  {\bf 60}, 569 (2010)
  [arXiv:1001.2799 [hep-ph]].

\bibitem{review2}
  A.~Mirizzi, I.~Tamborra, H.~T.~Janka, N.~Saviano, K.~Scholberg, R.~Bollig, L.~Hudepohl and S.~Chakraborty,
  ``Supernova Neutrinos: Production, Oscillations and Detection,''
  arXiv:1508.00785 [astro-ph.HE].

\bibitem{Kersten}
  J.~Kersten,
  ``Coherence of Supernova Neutrinos,''
  Nucl.\ Phys.\ Proc.\ Suppl.\  {\bf 237-238}, 342 (2013).

\bibitem{KerstSmir}
  J.~Kersten and A.~Y.~Smirnov,
  ``Decoherence and oscillations of supernova neutrinos,''
  arXiv:1512.09068 [hep-ph].

\bibitem{AM}
  E.~Akhmedov and A.~Mirizzi,
  ``Another look at synchronized neutrino oscillations,''
  Nucl.\ Phys.\ B {\bf 908} (2016) 382
  [arXiv:1601.07842 [hep-ph]].

\bibitem{AKL3}
  E.~Akhmedov, J.~Kopp and M.~Lindner,
  ``Decoherence by wave packet separation and collective neutrino oscillations,''
  arXiv:1405.7275 [hep-ph].

\bibitem{Kiers}
  K.~Kiers, S.~Nussinov and N.~Weiss,
  ``Coherence effects in neutrino oscillations,''
  Phys.\ Rev.\ D {\bf 53} (1996) 537
  doi:10.1103/PhysRevD.53.537
  [hep-ph/9506271].

\bibitem {Beuthe}
  M.~Beuthe,
  ``Oscillations of neutrinos and mesons in quantum field theory,''
  Phys.\ Rept.\  {\bf 375} (2003) 105
  [hep-ph/0109119].


\bibitem{parad}
  E.~K.~Akhmedov and A.~Y.~Smirnov,
  ``Paradoxes of neutrino oscillations,''
  Phys.\ Atom.\ Nucl.\  {\bf 72} (2009) 1363
  [arXiv:0905.1903 [hep-ph]].

\bibitem{JH}
H.-T.~Janka, W.~Hillebrandt,
``Monte Carlo simulations of neutrino transport in type II supernovae,''
Astronomy and Astrophysics Supplement Series {\bf 78} (1989) 375.

\bibitem{SR}
  G.~Sigl and G.~Raffelt,
  ``General kinetic description of relativistic mixed neutrinos,''
  Nucl.\ Phys.\ B {\bf 406} (1993) 423.

\bibitem{W}
  L.~Wolfenstein,
  ``Neutrino Oscillations in Matter,''
  Phys.\ Rev.\ D {\bf 17} (1978) 2369.

\bibitem{MS}
  S.~P.~Mikheev and A.~Y.~Smirnov,
  ``Resonance Amplification of Oscillations in Matter and Spectroscopy of
   Solar Neutrinos,''
  Sov.\ J.\ Nucl.\ Phys.\  {\bf 42} (1985) 913
   [Yad.\ Fiz.\  {\bf 42} (1985) 1441].

\bibitem{cardMSW}
  C.~Y.~Cardall and D.~J.~H.~Chung,
  ``The MSW effect in quantum field theory,''
  Phys.\ Rev.\ D {\bf 60} (1999) 073012
  [hep-ph/9904291].

\bibitem{AW}
  E.~K.~Akhmedov and A.~Wilhelm,
  ``Quantum field theoretic approach to neutrino oscillations in matter,''
  JHEP {\bf 1301} (2013) 165
  [arXiv:1205.6231 [hep-ph]].

\bibitem{pant1}
  J.~T.~Pantaleone,
  ``Neutrino flavor evolution near a supernova's core,''
  Phys.\ Lett.\ B {\bf 342} (1995) 250
  [astro-ph/9405008].

\bibitem{card1}
  C.~Y.~Cardall,
  ``Liouville equations for neutrino distribution matrices,''
  Phys.\ Rev.\ D {\bf 78} (2008) 085017
  [arXiv:0712.1188 [astro-ph]].

\bibitem{Giunti1}
  C.~Giunti,
  ``Coherence and wave packets in neutrino oscillations,''
  Found.\ Phys.\ Lett.\  {\bf 17} (2004) 103
  [hep-ph/0302026].

\bibitem{HS}
  D.~Hernandez and A.~Y.~Smirnov,
  ``Active to sterile neutrino oscillations: Coherence and MINOS results,''
  Phys.\ Lett.\ B {\bf 706} (2012) 360
  [arXiv:1105.5946 [hep-ph]].

\bibitem{AHS}
  E.~Akhmedov, D.~Hernandez and A.~Smirnov,
  ``Neutrino production coherence and oscillation experiments,''
  JHEP {\bf 1204} (2012) 052
  [arXiv:1201.4128 [hep-ph]].

\bibitem{MSm}
  H.~Minakata and A.~Y.~Smirnov,
  ``Neutrino Velocity and Neutrino Oscillations,''
  Phys.\ Rev.\ D {\bf 85} (2012) 113006
  [arXiv:1202.0953 [hep-ph]].

\bibitem{cristina1}
  S.~Galais, J.~Kneller and C.~Volpe,
  ``The neutrino-neutrino interaction effects in supernovae: the point
   of view from the matter basis,''
  J.\ Phys.\ G {\bf 39} (2012) 035201
  [arXiv:1102.1471 [astro-ph.SR]].

\bibitem{cristina2}
  S.~Galais and C.~Volpe,
  ``The neutrino spectral split in core-collapse supernovae: a magnetic resonance phenomenon,''
  Phys.\ Rev.\ D {\bf 84} (2011) 085005
  [arXiv:1103.5302 [astro-ph.SR]].

\bibitem{MikhSm2}
  S.~P.~Mikheyev and A.~Y.~Smirnov,
  ``Resonant neutrino oscillations in matter,''
  Prog.\ Part.\ Nucl.\ Phys.\  {\bf 23}, 41 (1989).


\bibitem{HansSmi}
  R.~S.~L.~Hansen and A.~Y.~Smirnov,
  ``The Liouville equation for flavour evolution of neutrinos and neutrino
   wave packets,''
  arXiv:1610.00910 [hep-ph].


\end{thebibliography}
\end{document}